# Dynamics of several two-level atoms in a one-mode cavity

Paolo Schwendimann and Antonio Quattropani

Institute of Theoretical Physics. Ecole Polytechnique Fédérale de Lausanne.

CH 1015 Lausanne-EPFL, Switzerland

The dynamics of few two-level atoms interacting with one cavity mode is described in master equation formalism. Two different configurations are considered: a transient one with all atoms initially in the excited state and a stationary one where a coherent pump acts on either the atoms or the cavity mode. Superradiance, resonant scattering and switching between a low and a high emission state are considered. Cooperation and decoherence as well as the statistics of the emitted radiation field in both cases are discussed.

PACS: 42.50.Ar, 42.50.Lc, 42.50.Pq

## I. Introduction

Cooperation and decoherence in atomic systems interacting with the radiation field have been intensively studied since the seminal paper by R. H. Dicke [1]. Starting from the results by Dicke [1], cooperative phenomena like superradiance [2], resonant scattering [3], and optical bistability [4, 5] have been addressed in the last decades. The cooperative effects resulting from the interaction between atoms and the radiation field were mainly discussed in systems containing a large number of atoms. The study of cooperation and of other characteristics of the radiation field-atoms interaction has experienced a revival in the last years in connection with the growing interest in subjects like the optical properties of Quantum Dots and Quantum Computing. Contrary to the large atom number systems considered in older literature, the systems discussed in these contexts mainly contain a small number of atoms and in limiting cases one atom only. On the other hand, the development of high quality semiconductor microcavities or photonic crystals provides cavities of a high quality, in which the one-mode approximation may be realized. Examples of systems showing the above characteristics and considered both in theory and in experiment in the last years are Quantum Dots embedded in microcavities [6-11], coupled cavities containing one or several atoms [12-18] and q-bits systems in the context of Quantum Computing [19, 20]. In particular, superradiance in a system consisting of two dye molecules placed in a Fabry-Pérot cavity has been experimentally achieved [21, 22]. An advantage of the few atoms advantage of the system considered here is that the dynamics and its stationary properties are described without introducing approximations involving dynamical variables. Therefore, a different description of several quantum-optical effects can be performed within this model and some approximations found in earlier literature can be tested. Working with a small number of atoms allows investigating cooperative effects and decoherence as well as quantum characteristics of the emission like sub-poissonian statistics, which are difficult to detect when the number of atoms become large. Finally in this context we are able to analyze in detail the interplay of the different coherent and incoherent interaction in the dynamics of the atoms-cavity system.

In this paper we start from a master equation description of the dynamics of a few, at most five, atoms interacting with one cavity mode and accounting for relaxation both of the atoms and of the radiation mode. We shall consider both the dynamics in absence of a coherent pump mechanism and the dynamics in presence of the coherent pump. The cooperative effects in a few-atoms system subject to incoherent pumping have been discussed in detail in two recent papers [23, 24]. In all but one of the examples to be discussed in the following, the atomic system is embedded in a cavity whose dimensions are smaller that the wavelength of the mode. This assumption is realistic in view of the small number of atoms considered. We also have chosen the relaxation rates and the coupling constants in such a way that strong atom-mode coupling is achieved. This condition is mostly satisfied in current systems of interest. The number of photonic states involved in the dynamics is restricted, and no other approximation is introduced in solving the dynamical system. This restriction is due to the fact that the time required by the numerical calculations rapidly grows with the number of particles considered. However, we control that for a fixed number of atoms augmenting the number of photonic states no relevant qualitative differences appear. Therefore, our results are exact in the space of states defined via this restriction. In the following we will discuss two different dynamical behaviors corresponding to different preparations of the initial states. In the first example, the atomic system is assumed to be in its most excited state and no photons are present in the cavity. Due to the choice of the initial state and to the presence of cavity losses, the stationary solution for the density operator is zero. This situation is reminiscent of the one characterizing superradiance [25, 26]. In earlier studies on superradiance the assumption of a bad cavity i.e. a cavity with large losses was introduced [25]. Within this assumption, the dynamics of the radiation field was assumed to follow adiabatically the one of the atomic system, which was then studied in detail. We reconsider the problem without any adiabatic approximation and generalizing the discussion to situations in which the relaxation rate of the cavity may be small and the adiabatic approximation is no more valid. Our aim is twofold: on one side we want to discuss in detail how decoherence acts on the cooperative behavior of the atomic system. Furthermore, we want to discuss how the quality of the cavity influences the characteristics of the atomic cooperation and of the emission statistics. In the second example we shall consider the dynamics of an atomic system interacting both with the cavity mode and with a coherent pump acting either on the atoms or on the cavity mode. When the pump acts on the atoms and in the bad cavity case, we reproduce the situation characterizing coherent resonant scattering and discuss it along the lines of the superradiant case. When the pump acts on the field mode, we reproduce the characteristics of a system of few atoms rapidly switching between a state of low and a state of strong emission. This behavior is reminiscent of the bistable behavior of a many-atom system. In our scheme we are able to discuss the dynamics of the intensity and of the atomic and mode correlation in detail. We discuss both the dynamics of the second order mode correlations and the spectra in the switching regime as well as in the regions of low and strong emission.

The paper is organized as follows: In Section II we discuss in detail the dynamics of the atoms-mode system accounting for different types of decoherence. In Section III based on the analysis performed in Section II the spontaneous affects in absence of a pump mechanism are discussed. In Section IV we discuss in detail the effects of coherent pump acting either on the atoms or on the field mode. Finally, some technical details concerning the dynamical equations are presented in the Appendix.

## II. Theory

**a) General Considerations**

We consider a one-mode cavity containing a small number $N$ of two-level atoms. The Hamiltonian for this system is

$$H = \hbar\omega_F a^+ a + \sum_{j=1}^{N} \hbar\omega_{A,j} \sigma_j^+ \sigma_j^- + \lambda\left(S^+ a + S^- a^+\right), \tag{1}$$

where

$$S^+ = \sum_{j=1}^{N}\left(\sigma_j^+\right) \text{ and } S^- = \sum_{j=1}^{N}\left(\sigma_j^-\right).$$

The atom-cavity dipole coupling is denoted by $\lambda$. The frequencies $\omega_{A,j}$ and $\omega_F$ are the frequencies of the free system. The collective spin operators $S^+, S^-$ obey angular momentum commutation rules. In Section IV, we shall introduce a pump, alternatively on the atoms or on the mode. When all atoms have the same energy i.e. $\hbar\omega_{A,i} = \hbar\omega_A$, the Hamiltonian (1) has two constants of motion

$$N_{TOT} = a^+ a + \sum_{j=1}^{N}\sigma_j^+ \sigma_j^-, \tag{2a}$$

$$\mathbf{S}^2 = S^+ S^- + S_3^2 - \hbar S_3, \tag{2b}$$

with

$$S_3 = \sum_{j=1}^{N}\sigma_{3j}. \tag{2c}$$

The evolution of the system is described by the master equation

$$i\hbar\frac{d\rho}{dt} = \left[H,\rho\right] + i\sum_{i=1}^{N}\left(\Lambda_F \rho + \Lambda_{A,i}\rho\right) = L\rho, \tag{3a}$$

where $H$ is the Hamiltonian (1), $L$ is the evolution super operator, and

$$\Lambda_F \rho = \kappa\left(\left[a\,\rho,a^+\right] + \left[a\,,\rho a^+\right]\right), \tag{3b}$$

$$\Lambda_{A,i}\rho = \gamma\left(\left[\sigma_i^- \rho,\sigma_i^+\right] + \left[\sigma_i^-,\rho\sigma_i^+\right]\right) \tag{3c}$$

describe the relaxation of mode and atoms respectively.

Without losses the dynamics of the system is described in the space spanned by the direct product of the n-photon states $|n\rangle$ and of the angular momentum states $|S,m\rangle$. These last states are eigenstates of the total spin operator squared $\mathbf{S}^2$ (2b) and of the z-component of the total spin $S_3$ (2c). The values of the quantities S and m are $0 \leq S \leq N/2$ and $-S \leq m \leq S$ respectively, $N$ being the number of atoms. Furthermore, when $\mathbf{S}^2$ is conserved, the selection rules $\Delta S = 0, \Delta m = \pm 1$ hold. When the cavity losses (3b) are introduced, $N_{TOT}$ is no more conserved and $\langle N_{TOT} \rangle$ evolves in time according to the equation

$$\frac{d}{dt}\langle N_{TOT} \rangle = Tr\rho(\Lambda_F N_{TOT}) = -2\kappa\langle a^+ a \rangle. \tag{4}$$

However, $\mathbf{S^2}$ is conserved allowing to describe the dynamics of the system with the same selection rules as in the case without losses. When the atomic energies are different from each other, the conservation of $\mathbf{S^2}$ is lost and $\langle \mathbf{S^2} \rangle$ evolves according to the equation

$$\frac{d}{dt}\langle \mathbf{S^2} \rangle = -i\sum_{i,j}(\omega_i - \omega_j)\langle \sigma_i^+ \sigma_j^- + \sigma_j^+ \sigma_i^- \rangle. \tag{5}$$

In this case the selection rules become $\Delta S = 0, \pm 1$ and $\Delta m = \pm 1$. We shall discuss the consequences of this non-conservation in the following. Finally, when the atomic relaxation is introduced $N_{TOT}$ and $\mathbf{S^2}$ are no more conserved also at resonance. The selection rules are the same as in the atomic non-resonant case. We discuss these generalizations of equations (4) and (5) at the end of the following subsection.

**b) An introductory example: two atoms in a cavity**

The dynamics of the system of $N$ atoms as described by Eq. (3) is in general discussed using numerical calculations and described in the space spanned by the product of photons and of single atom states. The numerical calculations, which we present in Sections III and IV are performed in this framework. However, in order to get a picture of the details of the dynamics, we may exploit the fact that there are the two constants of motion $N_{TOT}$ and $\mathbf{S^2}$ and describe the dynamics in the space spanned by the direct product of their eigenstates (collective atomic variables and n-photon states). Within this representation the effects of cooperation and decoherence are easily highlighted. An advantage of this representation is that some of the relevant dynamical features may already be discussed within a very simple system consisting of two atoms in a cavity. Therefore, in this subsection we set $N=2$ in the expression derived above. In a first step, we consider a situation in which the atoms are resonant with the cavity mode. In the two-atoms system considered, the total spin has the eigenvalues 1 and 0 identifying its triplet and singlet states. We assume that there is no pump and that the initial state of the system $|S=1, m=1, n=0\rangle$ contains two excited atoms and no photons. $N_{TOT}$ takes the values 0, 1, and 2. In order to discuss the dynamics of the system in detail, we have to start from the explicit form of the equations for the matrix elements of the density operator obtained from (3a). Without losses, the dynamics involves three multiplets according to $S=1$ and $N=0,1,2$. These multiplets are decupled and, once the initial condition is chosen, the dynamics happens inside of one specific multiplet. When the cavity relaxation is introduced, $N_{TOT}$ is not a constant of motion anymore. As a consequence eq. (4) applies. Indeed, the non-conservation of $N_{TOT}$ appears in (4) as a loss of energy from the system into the cavity mode. As a consequence the multiplets of $N_{TOT}$ are coupled to each other through the relaxation terms as shown in the following equations: for the multiplet $N_{TOT}=2$ we have

$$i\hbar \langle 1,1,0| \frac{d\rho}{dt} |1,1,0\rangle = \lambda\sqrt{2}\left(\langle 1,0,1|\rho|1,1,0\rangle - \langle 1,1,0|\rho|1,0,1\rangle\right), \tag{6a}$$

$$i\hbar \langle 1,1,0| \frac{d\rho}{dt} |1,0,1\rangle = \lambda\sqrt{2}\left(\langle 1,0,1|\rho|1,0,1\rangle - \langle 1,1,0|\rho|1,1,0\rangle\right) -$$
$$2\lambda\left(\langle 1,1,0|\rho|1,-1,2\rangle\right) - i\kappa \langle 1,1,0|\rho|1,0,1\rangle, \tag{6b}$$

$$i\hbar \langle 1,0,1| \frac{d\rho}{dt} |1,0,1\rangle = -\lambda\sqrt{2}\left(\langle 1,0,1|\rho|1,1,0\rangle - \langle 1,1,0|\rho|1,1,0\rangle\right) +$$
$$2\lambda\left(\langle 1,-1,2|\rho|1,0,1\rangle - \langle 1,0,1|\rho|1,-1,2\rangle\right) - 2i\kappa \langle 1,0,1|\rho|1,0,1\rangle, \tag{6c}$$

$$i\hbar \langle 1,0,1| \frac{d\rho}{dt} |1,-1,2\rangle = 2\lambda \left( \langle 1,-1,2|\rho|1,-1,2\rangle - \langle 1,0,1|\rho|1,0,1\rangle \right) -$$
$$3i\kappa \langle 1,1,0|\rho|1,-1,2\rangle, \tag{6d}$$

$$i\hbar \langle 1,-1,2| \frac{d\rho}{dt} |1,-1,2\rangle = 2\lambda \left( \langle 1,0,1|\rho|1,-1,2\rangle - \langle 1,-1,2|\rho|1,0,1\rangle \right) -$$
$$4i\kappa \langle 1,-1,2|\rho|1,-1,2\rangle. \tag{6e}$$

Notice that in (6), only at most two excitations are present in the system. The equations in the multiplet $N_{TOT}=1$ are

$$i\hbar \langle 1,0,0| \frac{d\rho}{dt} |1,0,0\rangle = \lambda \left( \langle 1,-1,1|\rho|1,0,0\rangle - \langle 1,0,0|\rho|1,-1,1\rangle \right) +$$
$$2i\kappa \langle 1,0,1|\rho|1,0,1\rangle \tag{7a}$$

$$i\hbar \langle 1,0,0| \frac{d\rho}{dt} |1,-1,1\rangle = \lambda \left( \langle 1,0,0|\rho|1,0,0\rangle - \langle 1,-1,1|\rho|1,-1,1\rangle \right) +$$
$$2\sqrt{2}i\kappa \langle 1,0,1|\rho|1,-1,2\rangle - i\kappa \langle 1,0,0|\rho|1,-1,1\rangle \tag{7b}$$

$$i\hbar \langle 1,-1,1| \frac{d\rho}{dt} |1,-1,1\rangle = -\lambda\sqrt{2} \left( \langle 1,0,0|\rho|1,-1,1\rangle - \langle 1,-1,1|\rho|1,0,0\rangle \right) +$$
$$4\kappa \langle 1,-1,2|\rho|1,-1,2\rangle - 2i\kappa \langle 1,-1,1|\rho|1,-1,1\rangle \tag{7c}$$

Finally, for the case $N_{TOT}=0$ we obtain

$$i\hbar \langle 1,-1,0| \frac{d\rho}{dt} |1,-1,0\rangle = 2i\kappa \langle 1,-1,1|\rho|1,-1,1\rangle. \tag{8}$$

In order to understand the details of the dynamics in presence of the reservoir, we rewrite the explicit form of the mode relaxation operator in (3b) as

$$\Lambda_F \rho = \kappa \left( 2a\rho a^+ - a^+ a \rho - \rho a^+ a \right). \tag{9}$$

Since the cavity relaxation operator $\Lambda_F \rho$ models the effects of the mirrors, the dissipation describes the photons' escape from the cavity (second and third term in (9)) and the emission of photons into the system corresponds to the reflection of photons into the cavity (first term in (9)). In the equations for the multiplets $N_{TOT}=1$ and $N_{TOT}=0$ a coupling different from zero to the $N_{TOT}=2$ and $N_{TOT}=1$ states respectively appears via contributions corresponding to emission of a photon from the reservoir into the system. These terms have an effect comparable to an absorption processes from $N_{TOT}=1$ to $N_{TOT}=2$ and from $N_{TOT}=0$ to $N_{TOT}=1$ respectively. Notice, that the coupling terms in the equations couple only matrix elements of the lower multiplet to the upper one. The contribution of the emission of a photon into the state with two excited atoms and no photon would have the form $2i\kappa\langle 1,1,1|\rho|1,1,1\rangle$ but it is missing in eq. (6a). As a consequence, the equations for the multiplet $N_{TOT}=2$ there is no coupling to the other two multiplets, the effect of the dissipative term in the master equation being of pure relaxation into the reservoir according to equation (9). This result is not astonishing because, for $t=0$ the value of $\langle N_{TOT}\rangle=2$ is fixed by the initial condition and $\langle N_{TOT}\rangle$ decays in time according to (4). Therefore, the number of excitations in the system (atoms or photons) is of at most two and the matrix element $\langle 1,1,1|\rho|1,1,1\rangle$ doesn't belong to the considered states space. For different values of the relaxation rate $\kappa$ the time evolution of atomic intensity $\langle S^+S^-\rangle$ in the multiplets behaves as shown in Figure 1.

**Figure 1. Time evolution of $\langle S^+S^-\rangle$ in the multiplets $N_{TOT}=1$ and $N_{TOT}=2$ in the resonant configuration, with $\gamma=0, \lambda=0.1,$ and for different values of the relaxation rate $\kappa$.**

a) $\kappa=0.1, N_{TOT}=2$, b) $\kappa=0.1, N_{TOT}=1$, c) $\kappa=0.001, N_{TOT}=2$,
d) $\kappa=0.001, N_{TOT}=1$.

For very small values of the relaxation we have a very small coupling between the multiplets. The contribution of the $N_{TOT}=2$ multiplet reproduces the oscillatory behavior of the perfect cavity case with a small damping, while the contribution of $N_{TOT} = 1$ is small due to the small transfer of energy between the multiplets. On the contrary, with growing values of the relaxation the oscillations in the contribution of $N_{TOT} = 2$ disappear and relaxation only is found. However, the contribution of $N_{TOT} = 1$ oscillates as a consequence of the more relevant absorption processes from the reservoir. These last oscillations disappear too for large relaxation rates $\kappa \geq 1$ as a consequence of the fast relaxation of the photon number. Out of resonance and for atoms with different frequencies, the conservation of the total spin doesn't hold anymore. In fact, the operator $\mathbf{S}^2$ commutes with the interaction Hamiltonian in (1) but doesn't commute with the atomic part of the free Hamiltonian i.e.

$$\left[\mathbf{S}^2, \sum_{j=1}^{2} \hbar\omega_i \sigma_j^+ \sigma_j^-\right] = \hbar(\omega_1 - \omega_2)\left(\sigma_1^+ \sigma_2^- + \sigma_2^+ \sigma_1^-\right). \tag{10}$$

As a consequence, $\langle \mathbf{S}^2 \rangle$ evolves according to the equation

$$\frac{d}{dt}\langle \mathbf{S}^2 \rangle = -iTr\rho\left(\Lambda_F \mathbf{S}^2\right) = -i(\omega_1 - \omega_2)\langle \sigma_1^+ \sigma_2^- + \sigma_2^+ \sigma_1^- \rangle. \tag{11}$$

Since $\mathbf{S}^2$ is not conserved, the selection rules are now $\Delta S = 0,1$ and $\Delta m = \pm 1$ thus allowing transition between the spin multiplets. This fact introduces some relevant changes in the equations describing the dynamics of the system. Due to the selection rule $\Delta S = 0,1$ transitions between the atomic triplet and singlet states are allowed. These transitions happen in absence of photons, because the operator $\mathbf{S}^2$ commutes with the interaction Hamiltonian Therefore, the dynamics between the $N_{TOT}$-multiplets is the same as in the previous case. The transitions between atomic triplet and singlet states originate in the free atomic Hamiltonian, which depends on the different atomic frequencies. Therefore the decoherence process induced by the non-conservation of $\mathbf{S}^2$ is intrinsic to the Hamiltonian (1) and is not due to any interaction with reservoirs. We show here only two of the modified equations describing the dynamics of a diagonal and of an off-diagonal matrix element of the density operator respectively. The full system of equation describing decoherence effects is found in the Appendix. The modified equations are

$$i\hbar \langle 1,1,0| \frac{d\rho}{dt} |1,0,1\rangle = \lambda\sqrt{2}\left(\langle 1,0,1|\rho|1,0,1\rangle - \langle 1,1,0|\rho|1,1,0\rangle\right) -$$
$$2\lambda\left(\langle 1,1,0|\rho|1,-1,2\rangle\right) - i\kappa\langle 1,1,0|\rho|1,0,1\rangle -$$
$$(\omega_1 + \omega_2)\langle 1,1,0|\rho|1,0,1\rangle/2 + (\omega_1 - \omega_2)\langle 1,1,0|\rho|0,0,1\rangle/2, \quad (12a)$$

$$i\hbar \langle 1,0,1| \frac{d\rho}{dt} |1,0,1\rangle = \lambda\sqrt{2}\left(\langle 1,1,0|\rho|1,0,1\rangle - \langle 1,0,1|\rho|1,1,0\rangle\right) +$$
$$2\lambda\left(\langle 1,-1,2|\rho|1,0,1\rangle - \langle 1,0,1|\rho|1,-1,2\rangle\right) -$$
$$2i\kappa\langle 1,0,1|\rho|1,0,1\rangle - 2i(\omega_1 - \omega_2)\operatorname{Im}\langle 1,1,0|\rho|0,0,1\rangle. \quad (12b)$$

The remaining equations have a similar form. In the equation for the diagonal matrix element a frequency-dependent relaxation term appears while in the equation for the off-diagonal matrix element a frequency dependent oscillatory term originates in the transitions between the atomic triplet and singlet states. Until now we have considered systems whose spatial dimensions are smaller than the mode wavelength. When this approximation is relaxed, the atomic positions appear in the interaction part of the Hamiltonian (1) in form of position-dependent phases and we have

$$H_{\text{int}} = \lambda\left(aS_k^+ + a^+ S_k^-\right),$$

where

$$S_k^\pm = \sum_{j=1}^{2} exp(\pm ikx_j)\sigma_j^\pm.$$

As a consequence, the total spin is not a constant of motion anymore because

$$[\mathbf{S}^2, H] = 2a(S^+ S_{zk+} - S_k^+ S_z) + 2a^+(S_{zk-}S^- - S_z S_k^-),$$

where

$$S_{zk\pm} = \sum_{j=1,2} \sigma_{zj} \exp(\pm ikx_j).$$

Like in the non-resonant case, the non-conservation of $\mathbf{S}^2$ implies that transition between the atomic triplet and singlet states appear in the dynamical equations.

Indeed, the time evolution of the diagonal matrix element of the master equation in the uppermost excited state and no photons reads

$$i\hbar \langle 1,1,0| \frac{d\rho}{dt} |1,1,0\rangle = \lambda\left(\exp(ikx_1) + \exp(-ikx_2)\right)\langle 1,0,1|\rho|1,1,0\rangle - c.c. + \lambda\left(\exp(ikx_1) - \exp(-ikx_2)\right)\langle 1,0,0|\rho|1,1,0\rangle - c.c.$$

showing that a phase-dependent coupling to the singlet atomic state appears in the equation. All the processes discussed above lead to a destruction of atomic coherence, which is not due to an interaction with a reservoir. Finally, a further violation of the conservation of $\mathbf{S}^2$ appears when the atomic relaxation is introduced. It is a consequence of the fact that the individual atoms are affected by the relaxation. Furthermore, the atomic relaxation also modifies the relaxation of $N_{TOT}$. The time evolution of both $N_{TOT}$ and $\mathbf{S}^2$ is expressed by

$$\frac{d}{dt}\langle N_{TOT}\rangle = -2\kappa\langle a^+ a\rangle - 4\gamma\left(\langle\sigma_1^+\sigma_1^-\rangle + \langle\sigma_2^+\sigma_2^-\rangle\right), \tag{13a}$$

$$\frac{d}{dt}\langle \mathbf{S}^2\rangle = -2\gamma\left(\langle\sigma_1^+\sigma_2^-\rangle + \langle\sigma_1^-\sigma_2^+\rangle + \langle\sigma_1^+\sigma_1^-\sigma_2^z\rangle + \langle\sigma_1^z\sigma_2^+\sigma_2^-\rangle\right). \tag{13b}$$

The atomic relaxation allows transitions between atomic singlet and triplet states as well as transitions between the $N_{TOT}$ multiplets analogous to the ones introduced by the cavity dissipation. However, in this case, these transitions involve the atomic states in the different multiplets only. The explicit form of the corresponding equations for the matrix elements of the density operator is found from the general equations in the Appendix when specified for the resonant case. The multiplets structure is independent of the number of atoms. Therefore the dynamics for $N > 2$ shows the same characteristics as that for $N = 2$.

## III. Cooperation and decoherence

In this Section we consider the case of a system of $N = 5$ two-level atoms embedded in a one-mode cavity. We chose as an initial state, a situation in which the system is prepared in its the upmost excited state without photons and with a very large atomic relaxation time such that we may assume $\gamma = 0$. Therefore the number of allowed photons is $N_{max} = 5$. In the framework of the description of Section IIb with $\gamma = 0$, the initial state is $|S = 5/2, m = 5/2, n = 0\rangle$ and the density operator at $t=0$ is a projector on this initial state. As a consequence, we shall describe transient effects induced by the spontaneous emission. The dynamics is described in the framework outlined in Section IIb with $\gamma = 0$. Therefore, when $\mathbf{S}^2$ is conserved the dynamics develops within the six multiplets corresponding to the values of $N_{TOT}$ coupled by the terms describing reflection of photons by the mirrors. When $\mathbf{S}^2$ is not conserved, the coupling to multiplets corresponding to values $S < 5/2$ also influence the dynamics, as outlined in Section IIb. Since without pumping the system is in a transient state, we consider the time-dependent behavior of the different characteristic quantities like intensity and correlations. In the following, we present results obtained from the numerical solution of the master equation (3). In order to better understand our results, we compare them for large $\kappa$ with the corresponding ones obtained from the well-established theory of superradiance. In this theory it is assumed that the radiation field instantaneously leaves the cavity allowing for the adiabatic approximation of the radiation field. In this situation $\mathbf{S}^2$ is conserved and cooperation between the atoms holds. Notice, that in this situation purely spontaneous effects govern the dynamics. Some important characteristics of the superradiant emission are the following: The emission has the form of a pulse whose intensity starts from spontaneous emission value proportional to $N$ and whose maximum is proportional to $N^2$. In Figure 2 we present the time-dependent behavior of the atomic intensity $\langle S^+S^-\rangle$ for large cavity decay $\kappa=1$, including the effects of decoherence, the result for superradiance is also indicated.

**Figure 2. Time dependence of the atomic intensity $\langle S^+S^-\rangle(t)$ with a cavity decay rate $\kappa=1$, a coupling $\lambda=0.1$, and no pump acting on the system.**

**a) resonant configuration $\omega_F = \omega_{A,i}$ and an atomic decay rate $\gamma=0.02$.**

**b) resonant configuration and $\gamma=0$.**

**c) superradiant result.**

**d) non-resonant configuration with the following detuning with respect to the cavity frequency $\omega_F$:**

$\Delta\omega_{A,1} = -0.01, \Delta\omega_{A,2} = 0.05, \Delta\omega_{A,3} = -0.07, \Delta\omega_{A,4} = 0.09, \Delta\omega_{A,5} = 0.03,$ **and $\gamma=0$.**

The behavior of $\langle S^+S^-\rangle$ both in the resonant configuration and with an atomic decay rate $\gamma=0$ almost coincides with the one obtained from superradiance calculated with the adiabatic approximation. When the atoms are out of resonance or $\gamma \neq 0$, we find some differences with respect to the superradiance case. In both cases the maximum of the intensity is reduced indicating that cooperation is less effective. Furthermore, the delay time needed to reach the maximum of the emission reduces with decoherence. However, the origin of this behavior is different in both cases. When $\gamma \neq 0$ the deviations from the superradiant behavior are due to the interaction with the atomic reservoir. On the contrary, in the non-resonant case the deviations from the superradiant behavior appear when $\gamma = 0$ and originate in the non-conservation of the total spin expressed by $\left[\mathbf{S}^2, \sum_{j=1}^{2} \hbar\omega_i \sigma_j^+ \sigma_j^-\right] \neq 0$. As a consequence, oscillations appear with growing time in the non-resonant case as expected from (5). This peculiar behavior of the intensity in the non-resonant case is better understood when looking for the time evolution of the correlation between two different atoms. It may be shown that in the superradiant case, the atom-atom correlation behaves like the intensity ensuring full cooperation. As a consequence, the phase difference between the atoms is zero. In the non-resonant case as well as when $\gamma \neq 0$, the real part of the correlation changes its sign indicating that there is a time-dependent difference of phase between the atoms. This phase-dependence is due to the fact that the dynamics of the individual atoms becomes relevant. It is of particular interest to understand how the decoherence mechanism acts in the atomic non-resonant case. In Figure 3 we present the evolution of the phase of the two-atom correlation for different values of the mode relaxation rate $\kappa$.

**Figure 3. Plot of the phase of $\langle \sigma_1^+ \sigma_5^- \rangle(t)$ as a function of time in a non-resonant configuration with the following detuning with respect to the cavity frequency $\omega_F$: $\Delta\omega_{A1} = -0.01$, $\Delta\omega_{A2} = 0.05$, $\Delta\omega_{A3} = -0.07$, $\Delta\omega_{A4} = 0.09$, $\Delta\omega_{A5} = 0.03$. The decay $\kappa$ takes various values, $\gamma = 0$, and $\lambda = 0.1$.**

In order to get an intuitive picture of what is happening, let us represent the two level atoms of the system by two spins ½. In the case of large mode relaxation $\kappa$ and at time $t=0$ the two spins are aligned in the up direction without any phase difference. With growing time, the phase difference becomes different from zero and grows corresponding to the second spin rotating away from the first one. When the phase difference becomes π the two spins look in opposite direction and, because of the intrinsic dephasing mechanism, the second spin begins to drift in the opposite direction. As a consequence, cooperation diminishes. Notice, that the decoherence mechanism behind this behavior originates in the non-conservation of **S²** and not in the interaction with the outside world. When the mode relaxation diminishes, the phase difference begins to oscillate as a consequence of the more rich dynamics between the atomic multiplets corresponding to different vales of *S*. We have also verified numerically that for $\kappa = 1$ the field intensity is proportional to the atomic one and fulfill the relation

$$\langle a^+ a \rangle = \lambda^2 \langle S^+ S^- \rangle / \kappa^2 \tag{14}$$

An analogous relation holds for the higher order moments as found in the theory of superradiance. For smaller value of $\kappa$, Eq. (14) doesn't hold anymore. Therefore it is of interest to discuss the statistics of the emitted photons for decreasing values of $\kappa$. In Figure 4 we present the time-dependent expectation value of the photon number $\langle a^+ a \rangle(t)$ as a function of time for the resonant case and without atomic relaxation.

**Figure 4. Time dependence of the average photon number $\langle a^+ a \rangle(t)$ for the resonant case $\omega_F = \omega_A$ and for various values of the cavity decay rate $\kappa$, with an atomic decay rate $\gamma = 0$ and a coupling $\lambda = 0.1$.**

When the relaxation rate of the mode is large, the radiation field rapidly vanishes from inside the cavity Therefore, the average photon number is very small when $\kappa = 1$ and the emitted photon number is proportional to the atomic intensity as already shown above. By increasing the quality of the mirrors, the relaxation diminishes and more photons remain inside the cavity leading to the absorption processes between multiplets analyzed in Section IIb. These processes lead to oscillations in the time dependence of the photon number. The oscillations that appear in Figure 4 represent the sum of the contributions of all multiplets presented in Figure 1. Furthermore, in this case, there is no proportionality relation between expectation values of atomic and of the mode operators. Therefore it is not possible to reproduce the photon number dynamics by simply scaling the atomic one. Finally, the cooperative behavior, which is effective in the build-up region of the evolution, is restricted to a smaller time interval. Indeed, as a consequence of the dynamics between multiplets, for lager time reabsorption and reemission effects balance spontaneous effects. The effects of the photon exchanges between the cavity mode and the reservoir appear in the statistical properties of the emitted radiation field too. In particular, we consider the quantity $g^2(t) - 1 = \langle a^+ a^+ aa \rangle(t) / \langle a^+ a \rangle^2(t) - 1$ expressing the deviation from coherence of the second moment of time-dependent photon number distribution and the two-times-dependent normalized second order correlation $g^{(2)}(t_1, t_2) = \langle a^+(t_1) a^+(t_2) a(t_2) a(t_1) \rangle / \langle a^+(t_2) a(t_2) \rangle \langle a^+(t_1) a(t_1) \rangle$. This last quantity allows studying the dependence of the correlation time on the radiation damping. In Figure 5 we present the photon second order correlation.

**Figure 5. Plot of the photon second order correlation $g^{(2)}(t) - 1$ as a function of time for various values of $\kappa$, $\gamma = 0$, and $\lambda = 0.1$.**

The quantity $g^{(2)}(t)-1$ plotted on Figure 5 is a measure of the deviation of the time-dependent statistics of the photon field from the one of a coherent state. In the bad cavity case ($\kappa = 1$) the statistics at t=0 is characteristic of an incoherent field. In course of time the difference between the coherent and the incoherent behavior diminishes and the emitted field becomes nearly coherent when the maximum of the emitted intensity is achieved corresponding to maximum atomic cooperation. For larger times the emission becomes once more incoherent since the cooperation is reduced in time according to Figure 2. When $\kappa$ is reduced, the minima in the correlations shift to earlier times reflecting the fact that cooperation happens on a shorter time scale. The fluctuations are strongly reduced and at the point where the intensity is maximal they have a negative value. The negative minima indicate that the emitted radiation has the behavior characteristic of sub-poissonian statistics in the time domain. Finally, the oscillations appearing for $\kappa = 0.1$ are a consequence of the oscillatory character of $\langle a^+ a \rangle(t)$ and $\langle a^+ a^+ aa \rangle(t)$ magnified when taking their ratio.

In the non-resonant case, the fluctuations show characteristics analogous to the ones in Figure 5 and their behavior has a similar interpretation. However, the fluctuations in the region of maximum emission are larger than in the resonant case preventing to achieve coherent statistics. The subpoissonian characteristic found in Figure 5 is also strongly reduced for $\kappa = 0.1$. In fact, this case we are in presence both of the intrinsic fluctuations of the cooperative decay and of the decoherence effects characteristic of the non-resonant case. Some more results concern the two-times correlation function for different values of $\kappa$, from which the cooperation time is obtained are presented in Figure 6.

**Figure 6. Plot of the two-time-dependent normalized second order correlation** $g^{(2)}(t_1,t_2)$ **for** $t_1 = 2$ **units and** $t_2 \geq t_1$ **in the resonant configuration, for various values of** $\kappa$**, for** $\gamma = 0$**, and** $\lambda = 0.1$**.**

Since we are studying a transient effect with vanishing stationary state, the two-time correlation shows the characteristics of decay. The relaxation of the correlation is not an exponential as expected from the dynamics of the atomic system and shows an oscillating behavior for small relaxation. The peculiar behavior of the correlation for larger time has the same origin as in Figure 5. The cooperation time is strongly reduced as a function of the mode decay rate.

## IV. Dynamics in the presence of a pump.

In this Section, we discuss the behavior of the system introduced in Sect. II when a pump acts alternatively on the atoms or on the cavity mode. In the following, we consider the resonant case. The action of the pump on the atoms and on the field modes respectively is described by

$$H_{pump,A} = \sum_{j=1}^{N} \eta_j \left( \sigma_j^- \exp(-i\omega_p t) + \sigma_j^+ \exp(i\omega_p t) \right) + $$
$$\alpha \left( a \exp(-i\omega_p t) + a^+ \exp(i\omega_p t) \right). \quad (15)$$

The time dependence of the pump is eliminated from the Hamiltonians through the unitary transformation

$$Z(t) = \exp\left[ \sum_{j=1}^{N} \left( \sigma_j^+ \sigma_j^- + a^+ a \right) i\omega_p t \right]. \quad (16)$$

As a consequence, he Hamiltonian describing the couplings in this model reads

$$H = H_0 + \lambda\left(S^+ a + S^- a^+\right) + \eta\sum_{j=1}^{N}\left(\sigma_j^- + \sigma_j^+\right) + \alpha\left(a + a^+\right), \quad (17a)$$

$$H_0 = \sum_{1}^{N} \hbar\omega_A \sigma_j^+ \sigma_j^- + \hbar\omega_F a^+ a. \quad (17b)$$

Here $\eta$ and $\alpha$ are the amplitudes of the pumps acting on the atoms and on the modes, respectively. The frequencies $\omega_A$ and $\omega_F$ in $H_0$ are the frequencies of the free system shifted by the pump frequency $\omega_p$ and measured in units of the pump frequency $\omega_p$. In the following, we shall consider the pumping mechanisms separately by choosing either $\eta = 0$ (pump on the cavity mode) or $\alpha = 0$ (pump on the atoms). The presence of the pump introduces relevant differences within the dynamics of the system with respect to the results in Sections IIb and III. First of all the quantity $N_{TOT}$ is not conserved also when $\kappa = 0$. Indeed, when the pump acts on the cavity mode the time evolution reads

$$\frac{d\langle N_{Tot}\rangle}{dt} = -\kappa\langle N_{Tot}\rangle - i\alpha\left(\langle a^+\rangle - \langle a\rangle\right),$$

while when the pump acts on the atoms one has

$$\frac{d\langle N_{Tot}\rangle}{dt} = -\kappa\langle N_{Tot}\rangle - i\eta\left(\langle S^+\rangle - \langle S^-\rangle\right).$$

Therefore, when $\alpha = 0$ or $\eta = 0$ the interaction with the pump acting on the atoms or on the mode, respectively introduces direct transitions between the multiplets discussed in Section II complicating the picture of the dynamics. However, there is conservation of $\mathbf{S}^2$ when $\gamma = 0$ thus ensuring a degree of cooperation. Finally, the dynamics tends to a stationary regime where the density matrix is different from zero allowing the calculation of the stationary time-dependent correlations as well as of the stationary spectra of the radiation field. In the following we assume that in the initial state the system is in the ground state and the influence of the pump becomes manifest for times $t > 0$. It is evident that the stationary state of the system will be a mixed atoms-photon state. We expect the system to behave differently depending on the pump configuration. When the pump acts on the atoms, the atomic excitation saturates for high values of the pump as a consequence of the finite number of atomic levels considered. The photon number will decrease more or less rapidly depending on the magnitude of the cavity relaxation rate. When the pump acts on the field mode, the saturation mechanism for the atoms remains the same as in the former case. However, the photon number will grow further on due to the presence of the pump on the mode. In the following, we shall discuss these two regimes in detail.

a) **The pump acts on the atoms**

When the pump acts on the atoms, the stationary solutions of both atoms and photon number behave as shown in Figure 7.

**Figure 7. Plot of the atomic intensity $\langle S^+ S^- \rangle(t)$ and field intensity $\langle a^+ a \rangle(t)$ as a function of time for five 2-level atoms in a one-mode cavity, $N_{max} = 5$. At time t=0 the system is in the ground state. The parameters are $\kappa = 1$, $\gamma = 0$, and $\lambda = 0.1$, with a pump intensity $\eta = 0.04$ acting on the atoms.**

In Figure 7, as well as in Figures 8 and 9, the number on considered photons $N_{max} = 5$. In the case of only one atom interacting with the cavity mode and for a large cavity relaxation rate, antibunching is expected in the normalized stationary second order correlation functions. These for the radiation field and for the atomic quantities are defined as

$$g_F^{(2)}(t) = \langle a^+(0) a^+(t) a(t) a(0) \rangle_{stat} / \langle a^+(0) a(0) \rangle_{stat}^2, \tag{18a}$$

$$g_A^{(2)}(t) = \langle S^+(0) S^+(t) S^-(t) S^-(0) \rangle_{stat} / \langle S^+(0) S^-(0) \rangle_{stat}^2. \tag{18b}$$

Here the average is taken using the stationary density matrix for the system. In general, antibunching occurs when $g^{(2)}(t) > g^{(2)}(0)$. Notice that on a sufficiently long time scale $g^{(2)}(t) \to 1$. Thus a situation for which $g^{(2)}(t) < 1$ will always exhibit antibunching on some time scale. The amount of antibunching is known to decrease when the number of atoms in the system is increased. We verify this behavior in Figure 8, where we plot the time evolution of $g_F^{(2)}(t)$ and $g_A^{(2)}(t)$.

**Figure 8. Plot of $g_F^{(2)}(t)$ and $g_A^{(2)}(t)$ as a function of time for various number of atoms in a one mode bad cavity. The parameters are $\kappa = 1$, $\gamma = 0.01$, $\lambda = 0.1$ with a pump intensity $\eta = 0.04$ acting on the atoms.**

The result for one atom corresponds to the situation in resonant fluorescence and coincides with the one by Walls [27]. With growing atoms number the antibunching effects diminishes and disappears when $N = 4$. These results correspond to the ones found in the literature. Note that there is no significant difference between the results for $g_F^{(2)}(t)$ and $g_A^{(2)}(t)$. We shall come back to this point later on. With growing pump intensity the antibunching effect disappears for $N \geq 2$. This behavior is understood as follows: Antibunching in the one-atom case appears because the atom after the emission of one photon has to go back to the excited state before a second emission takes place. When more than one atom is present, the inter-atomic cooperation complicates the emission process. When the pump is increased the atoms have a larger probability of being in the excited state and the transition probability to the ground state of the system decreases, contrasting the development of antibunching. When the cavity quality improves, the mode radiation rate diminishes and this fact has consequences concerning the behavior of the correlations as shown in Figure 9.

**Figure 9. Plot of $g_F^{(2)}(t)$ and $g_A^{(2)}(t)$ as a function of time for five 2-level atoms in a one-mode cavity, $N_{max} = 5$. The parameters are $\gamma =0.01$, $\lambda = 0.1$ with a pump intensity $\eta =0.04$ acting on the atoms. The correlation $g_F^{(2)}(t)$ for $\kappa =0.01$ shows very small oscillations around the value 1 and is not reported here.**

The correlations of the atoms and of the radiation mode respectively become different for short times. As a consequence, a small sub-poissonian characteristic is found in the field correlation in contrast to the bad cavity case. Notice that the atoms correlation doesn't show any antibunching effect. For larger times both correlations show a coherent behavior as expected. Contrary to the bad cavity case, in a good cavity the sub-poissonian characteristic for short times for the photon correlation is enhanced when the pump intensity grows. The mechanism behind this behavior is the same, which is responsible for the sub-poissonian characteristics of the emission in the good cavity case in the former Section (Figure 5). The difference between atoms and mode correlations shown in Figure 9 are understood as follows: The statistics of the radiation filed mode can be also calculated in the equivalent Langevin description. The Langevin equations for both the field mode amplitude operator and the atomic variable $S^-$ are

$$\frac{da}{dt} = -i\lambda S^- - \kappa a + F_a(t), \tag{19}$$

$$\frac{dS^-}{dt} = -2i\lambda a S_z - \gamma S^- - 2i\eta S_z + F_{S^-}(t), \tag{20}$$

where $F_a(t)$ and $F_{S^-}(t)$ are the usual Langevin operators.

For large mode relaxation as it is the case in the bad cavity regime, the time derivative of the mode amplitude is neglected obtaining

$$\kappa a = -i\lambda S^- + F_a(t). \tag{21}$$

It follows from (21) that the expectation values of the mode and the atomic operators are proportional to each other. Therefore, when calculating mode correlations, these will be proportional to the atomic ones and are equal when normalized as already seen in Section III, Eq. (14). This justifies the results in Figure 8. For smaller radiation rate, the derivative in (19) cannot be neglected leading to a difference between the two correlations as shown in Figure 9. Finally, some relevant information on the characteristics of the stationary state is found when considering the spectrum of the mode i.e. the time Fourier transform of its two-operator time-dependent correlation

$$g_F^{(1)}(t) = \langle a^+(t) a(0) \rangle_{stat} + \langle a^+(0) a(t) \rangle_{stat}. \tag{22}$$

In Figure 10, for numerical reasons we present the spectrum of a simpler system in which only two atoms are present and $N_{max} = 4$. In fact, augmenting the number of atoms doesn't change the main structure of the spectrum.

**Figure 10. Plot of the photon spectrum $g_F^{(1)}(\omega)$ as a function of the frequency $\omega$ for two 2-level atoms in a one-mode cavity, $N_{max} = 4$ for $\kappa = 1$, $\gamma = 0.01$, $\lambda = 0.1$, and for various values of the pump intensity $\eta$ acting on the atoms.**

For small pump the spectrum consists of a doublet of very large lines centered on the frequency corresponding to the eigenvalues of the Hamiltonian with the pump on the atom (Eq. 16). For growing pump values a sharp peak develops at zero frequency as a consequence of the presence of the pump. This spectral structure is characteristic of a Mollow triplet [28]. Furthermore, the maxima of the sidebands change their position as a consequence of the dynamical Stark effect induced by the pump. All these characteristics confirm that in the stationary state of the system is in a mixed field-atoms state.

**b) Pump acting on the mode**

When the pump acts on the field mode, the emission characteristics of the system are quite different from the ones discussed above in the pump-on-the atom case. This fact is clearly shown when writing the Langevin equations for the mode amplitude and the polarization operators analogous to (19) and (20)

$$\frac{da}{dt} = -i\lambda S^- - \kappa a - i\alpha + F_a(t), \qquad (23)$$

$$\frac{dS^-}{dt} = -2i\lambda a S_z - \gamma S^- + F_{S^-}(t), \qquad (24)$$

which in the bad cavity case take the form

$$\kappa a = -i\lambda S^- - i\alpha + F_a(t). \qquad (25)$$

The pump amplitude $\alpha$ appears now in the equation for the mode. When the atomic system saturates under the influence of the pump, the photon number will continue to grow with $\alpha$. The behavior of the number of photons in the mode in the stationary regime in function of the pump intensity is shown in Figure 11 for very small cavity losses for $N = 4$.

**Figure 11. Plot in the resonant configuration of the number of photons $\langle a^+ a \rangle_{stat.}$ in the cavity mode and of the atomic intensity $\langle S^+ S^- \rangle_{stat.}$ in the stationary regime**

**as a function of the pump intensity $\alpha$ acting on the field mode. Four 2-level atoms are embedded in an unique cavity, $N_{max} = 7$. The cavity decay rate $\kappa = 0.01$, the atomic decay rate $\gamma = 0.01$ and the coupling $\lambda = 0.1$.**

Inside a small interval of pump values, the system switches from a state low photon emission into a state of high emission indicating the presence of a threshold. This behavior is reminiscent of the one characterizing optical bistability. Indeed, the behavior shown in Figure 11 is analogous to the one found in the literature [5] for a system consisting of a large number of atoms. Notice that our result is obtained in a scheme in which the only approximation consists in restricting the number of photon states as already stated in the introduction. The threshold is more and more pronounced for increasing number of photon states included in the evaluation, but the qualitative behavior doesn't change. We have also verified that the threshold for the transition to strong emission grows with growing mode relaxation $\kappa$ and the transition disappears when $\kappa > 0.1$. Also notice, that there is a transition region corresponding to the interval of pump values from the threshold value up to the point at which the high emission region sets in. As indicated in Figure 11, the threshold occurs in the interval $0.06 < \text{pump} < 0.27$ for the choice of the system parameters indicated in the caption. The characteristics of the transition between the emission states are investigated by considering the behavior of the fluctuations of the mode presented in Figure 12 and the spectrum of the emission presented in Figure 13.

**Figure 12. Plot in the resonant configuration of the photon stationary fluctuations $\langle a^+(0)a^+(t_0)a(t_0)a(0)\rangle_{stat.} - \langle a^+a\rangle^2_{stat.}$ for various fixed times $t_0$ as a**

**function of the pump intensity $\alpha$ acting on the field mode. Four two-level systems are embedded in a unique cavity, $N_{max} = 7$. The cavity decay rate $\kappa = 0.01$, the atomic decay rate $\gamma = 0.01$ and the coupling $\lambda = 0.1$.**

As shown in Figure 12, in the regimes of weak and strong emission, the fluctuations for long times become very small indicating that the mode statistics is rapidly approaching the one of a coherent state. However, in the transition region the fluctuations are large indicating that the collective behavior of the system undergoes a dramatic change. An analogous behavior is found when considering the atomic correlations. We have also verified that in the transition region relevant atom-atom correlations develop during the time evolution in analogy with the results found for cooperative spontaneous emission in Sect. III. Finally, relevant atom-field mode correlations also develop in the transition region. All these results indicate that, in the transition region, cooperative processes induced by the field mode determine the behavior of the system. When going over to the strong emission region the interaction with the pump becomes dominant. Analogous information is obtained from the spectrum of the field mode as indicated in Figure 13.

**Figure 13. Plot in the resonant configuration of the photon spectrum $g_F^{(1)}(\omega)$ as a function of the frequency $\omega$ for $\kappa = \gamma = 0.01$, $\lambda = 0.1$, and for various values of the pump intensity $\alpha$ acting on the field mode. Four two-level systems are embedded in a unique cavity, $N_{max} = 7$.**

Two interactions affect the behavior of the atoms: the interaction with the photons emitted by the atoms and the interaction with the pump. Well below the transition threshold i.e. for small pumping the spectrum consists of two lines, which originate in the atoms-mode interaction. When the pump amplitude is increased until $\alpha \sim 0.1$ a third line at $\omega = 0$ develops as a consequence of the effect of the pump. In this pump region the behavior of the spectrum is similar to the one found in the one atom case. This fact indicates that atomic cooperation and atom-mode correlation are negligible. In the transition region ($0.1 < \alpha < 0.21$) the spectrum becomes more involved: the atomic lines split into a doublet one of whose components shifts towards the central line with growing pump and disappears when $\alpha \sim 0.21$. This behavior is a consequence of the interplay between atomic cooperation and strong atom-mode correlations. At the same time, the width of the central line grows showing too that atomic cooperation, leading to larger relaxation rates, becomes relevant. For pump values $0.22 < \alpha < 0.3$ the central line develops sidebands leading to a Mollow triplet once more in analogy to the one atom case indicating that the effect of cooperation becomes smaller and the same holds for the linewidth of the central line. This result corresponds to the one found in the literature [3, 5]. For larger values of the pump the Mollow triplet disappears and the linewidth of the central line saturates indicating that no more cooperation is active. Indeed in this situation the atomic second order correlation shows a coherent behavior for long times. As a consequence, the behavior of the spectrum is analogous to the one found in the one-atom case.

## V. Conclusions

In this paper we have addressed the description of the atomic and radiation field dynamics of a few atoms embedded in a one-mode cavity in the framework of the master equation. The only approximation that we have introduced consists in a truncation of the state space of the photons. Within this truncation our results are exact. Therefore we have been able discuss the dynamics both of the atomic system and of the cavity mode on the same footing avoiding for instance the adiabatic approximation. Within this scheme we have discussed both the emission from a fully excited atom system in absence of photons and the effect of a coherent pump acting alternatively on the cavity mode or on the atoms. Besides reproducing the characteristics of superradiance and resonant scattering in the bad cavity limit as well as a switching analogous to the optical bistability in the corresponding limit, we discuss the atomic and photon correlations for different cavity loss regimes and highlight the effects of atomic cooperation in the different regimes considered. In analogy with the one atom case, we expect that considering few atoms in coupled cavities shall lead to some interesting results in function of the cavity coupling. In particular, we expect that when pumping one cavity only, the presence of the second cavity shall modify the characteristics of the effects discussed in this paper for the one mode case. The detailed discussion of this configuration is the goal of future investigations.

**Appendix**

We present the equations describing the evolution of the matrix elements of the density operator according to the master equation (3a) in the non-resonant case and with both atomic and mode relaxation. The matrix elements are calculated using the direct product of the n-photon states $|n\rangle$ and of the angular momentum states $|S,m\rangle$.

These states are eigenstates of the total spin operator squared $\mathbf{S}^2$ (2b) and of the z-component of the total spin $S_z$ (2c). In the present context it holds $0 \leq S \leq N/2$ and $-S \leq m \leq S$, $N$ being the number of atoms.

We consider a system consisting of two atoms interacting with a cavity mode. In this case $S = 1, 0$ and $m$ has the values 1, 0 and -1. The state space of this system is organized in multiplets corresponding to the three different values of the total number of excitations $N1$ (2a).

For the triplet $N1 = 2$ the five non trivial equations for the matrix elements of the density matrix read

$$i\hbar \langle 1,1,0| \frac{d\rho}{dt} |1,1,0\rangle = \lambda\sqrt{2}\left(\langle 1,0,1|\rho|1,1,0\rangle - \langle 1,1,0|\rho|1,0,1\rangle\right) - 4i\gamma \langle 1,1,0|\rho|1,1,0\rangle \quad (A1a)$$

$$i\hbar \langle 1,1,0| \frac{d\rho}{dt} |1,0,1\rangle = \lambda\sqrt{2}\left(\langle 1,0,1|\rho|1,0,1\rangle - \langle 1,1,0|\rho|1,1,0\rangle\right) - $$
$$2\lambda\left(\langle 1,1,0|\rho|1,-1,2\rangle\right) - i\kappa \langle 1,1,0|\rho 1,|0,1\rangle - 2i\gamma \langle 1,1,0|\rho|1,0,1\rangle - $$
$$(\omega_1 + \omega_2)\langle 1,1,0|\rho|1,0,1\rangle/2 + (\omega_1 - \omega_2)\langle 1,1,0|\rho|0,0,1\rangle/2 \quad (A1b)$$

$$i\hbar \langle 1,0,1| \frac{d\rho}{dt} |1,0,1\rangle = -\lambda\sqrt{2}\left(\langle 1,0,1|\rho|1,1,0\rangle - \langle 1,1,0|\rho|1,0,1\rangle\right) - 4i\gamma \langle 1,0,1|\rho|1,0,1\rangle + $$
$$2\lambda\left(\langle 1,-1,2|\rho|1,0,1\rangle - \langle 1,0,1|\rho|1,-1,2\rangle\right) - $$
$$2i\kappa \langle 1,0,1|\rho|1,0,1\rangle - 2i(\omega_1 - \omega_2)\operatorname{Im}\langle 1,1,0|\rho|0,0,1\rangle \quad (A1c)$$

$$i\hbar \langle 1,0,1| \frac{d\rho}{dt} |1,-1,2\rangle = 2\lambda\left(\langle 1,-1,2|\rho|1,-1,2\rangle - \langle 1,0,1|\rho|1,0,1\rangle\right) - $$
$$3i\kappa \langle 1,1,0|\rho|1,-1,2\rangle - 2i\gamma \langle 1,0,1|\rho|1,-1,2\rangle + $$
$$(\omega_1 + \omega_2)\langle 1,0,1|\rho|1,-1,2\rangle/2 - (\omega_1 - \omega_2)\langle 0,0,1|\rho|1,-1,2\rangle/2 \quad (A1d)$$

$$i\hbar \langle 1,-1,2| \frac{d\rho}{dt} |1,-1,2\rangle = 2\lambda\left(\langle 1,0,1|\rho|1,-1,2\rangle - \langle 1,-1,2|\rho|1,0,1\rangle\right) - $$
$$4i\kappa \langle 1,-1,2|\rho|1,-1,2\rangle - 4i\gamma \langle 1,-1,2|\rho|1,-1,2\rangle \quad (A1e)$$

Equations for the transitions inside the doublet states $|1,-1,1\rangle$ and $|1,0,0\rangle$

$$i\hbar \langle 1,0,0|\frac{d\rho}{dt}|1,0,0\rangle = \lambda\left(\langle 1,-1,1|\rho|1,0,0\rangle - \langle 1,0,0|\rho|1,-1,1\rangle\right) + 2i\kappa \langle 1,0,1|\rho|1,0,1\rangle +$$
$$2i\gamma\langle 1,1,0|\rho|1,1,0\rangle - 4i\gamma\langle 1,0,0|\rho|1,0,0\rangle -$$
$$2i(\omega_1 - \omega_2)\text{Im}\langle 1,1,0|\rho|0,0,0\rangle \qquad (A2a)$$

$$i\hbar \langle 1,0,0|\frac{d\rho}{dt}|1,-1,1\rangle = \lambda\left(\langle 1,0,0|\rho|1,0,0\rangle - \langle 1,-1,1|\rho|1,-1,1\rangle\right) -$$
$$2i\gamma\langle 1,0,0|\rho|1,-1,1\rangle + 2i\kappa\langle 1,0,1|\rho|1,0,1\rangle - i\kappa\langle 1,0,0|\rho|1,-1,1\rangle -$$
$$(\omega_1 - \omega_2)\langle 0,0,0|\rho|1-1,1\rangle/2 + (\omega_1 + \omega_2)\langle 1,0,0|\rho|1,-1,1\rangle/2 \qquad (A2b)$$

$$i\hbar\langle 1,-1,1|\frac{d\rho}{dt}|1,-1,1\rangle = -\lambda\sqrt{2}\left(\langle 1,0,0|\rho|1,-1,1\rangle - \langle 1,-1,1|\rho|1,0,0\rangle\right) +$$
$$4\kappa\langle 1,-1,2|\rho|1,-1,2\rangle - 2i\kappa\langle 1,-1,1|\rho|1,-1,1\rangle + 2i\gamma\langle 1,0,1|\rho|1,0,1\rangle \qquad (A2c)$$

Equations for the fundamental state $|1,-1,0\rangle$

$$i\hbar\langle 1,-1,0|\frac{d\rho}{dx}|1,-1,0\rangle = 2i\kappa\langle 1,-1,0|\rho|1,-1,1\rangle + 4i\gamma\langle 0,0,0|\rho|0,0,0\rangle \qquad (A3)$$

Equations for the transition between atomic triplet and singlet states

$$i\hbar\langle 1,0,0|\frac{d\rho}{dt}|0,0,1\rangle = -i(\kappa + 2\gamma)\langle 1,0,0|\rho|0,0,1\rangle -$$
$$(\omega_1 - \omega_2)\langle 1,0,0|\rho|1,0,1\rangle/2 + (\omega_1 + \omega_2)\langle 1,0,0|\rho|0,0,1\rangle/2 \qquad (A4a)$$

$$i\hbar\langle 0,0,1|\frac{d\rho}{dt}|1,-1,2\rangle = -3i\kappa\langle 0,0,1|\rho|1,-1,2\rangle -$$
$$i2\gamma\langle 0,0,1|\rho|1,-1,2\rangle - (\omega_1 - \omega_2)\langle 1,0,1|\rho|1,-1,2\rangle/2 +$$
$$(\omega_1 + \omega_2)\langle 0,0,1|\rho|1,-1,0\rangle/2 \qquad (A4b)$$

$$i\hbar\langle 0,0,0|\frac{d\rho}{dt}|1,-1,1\rangle = -i(\kappa + 2\gamma)\langle 0,0,0|\rho|1-1,1\rangle -$$
$$(\omega_1 - \omega_2)\langle 1,0,0|\rho|1,-1,1\rangle/2 + (\omega_1 + \omega_2)\langle 0,0,0|\rho|1,-1,1\rangle/2 \qquad (A4c)$$

Equations for the diagonal evolution of the atomic singlet-photon states

$$\frac{d}{dt}\langle 0,0,0|\rho|0,0,0\rangle = -4\gamma\langle 0,0,0|\rho|0,0,0\rangle + 4\gamma\langle 1,1,0|\rho|1,1,0\rangle \qquad (A5a)$$

$$\frac{d}{dt}\langle 0,0,1|\rho|0,0,1\rangle = -4\gamma\langle 0,0,1|\rho|0,0,1\rangle \qquad (A5b)$$

We notice that the atomic relaxation is responsible for two modifications in the equations for the matrix elements of the density operator. The first modification is related to the non-conservation of the total angular momentum and introduces transitions between singlet and triplet states containing zero or one photon. The second modification is that the atomic relaxation term in (3c) leads to coupling between the $N_{TOT}$ multiplets having the same characteristics as the ones already discussed for the cavity relaxation. Therefore we have besides the dissipation, transitions to multiplets with a higher value of $N_{TOT}$ for values of $N_{TOT} < 2$.

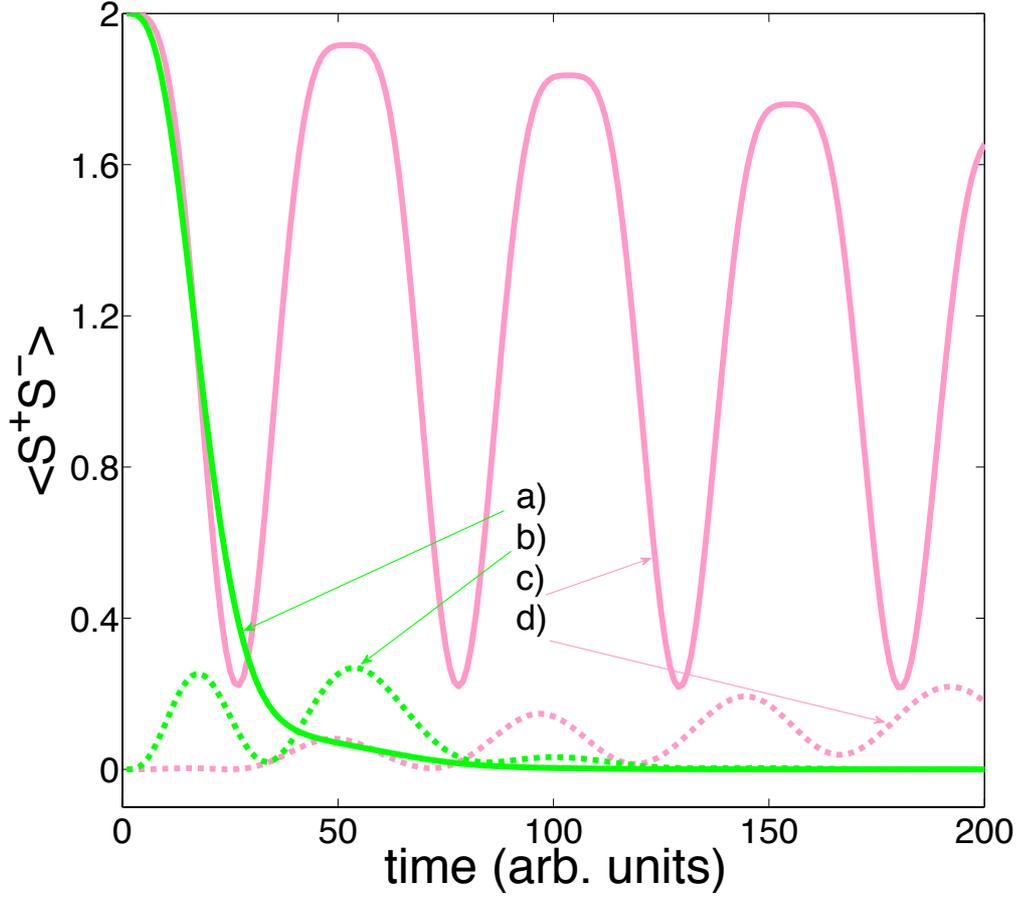

**Figure 1. Time evolution of $\langle S^+ S^- \rangle$ in the multiplets $N_{TOT}$=1 and $N_{TOT}$=2 in the resonant configuration, with $\gamma = 0, \lambda = 0.1,$ and for different values of the relaxation rate $\kappa$.**
a) $\kappa = 0.1, N_{TOT} = 2$, b) $\kappa = 0.1, N_{TOT} = 1$, c) $\kappa = 0.001, N_{TOT} = 2$,
d) $\kappa = 0.001, N_{TOT} = 1$.

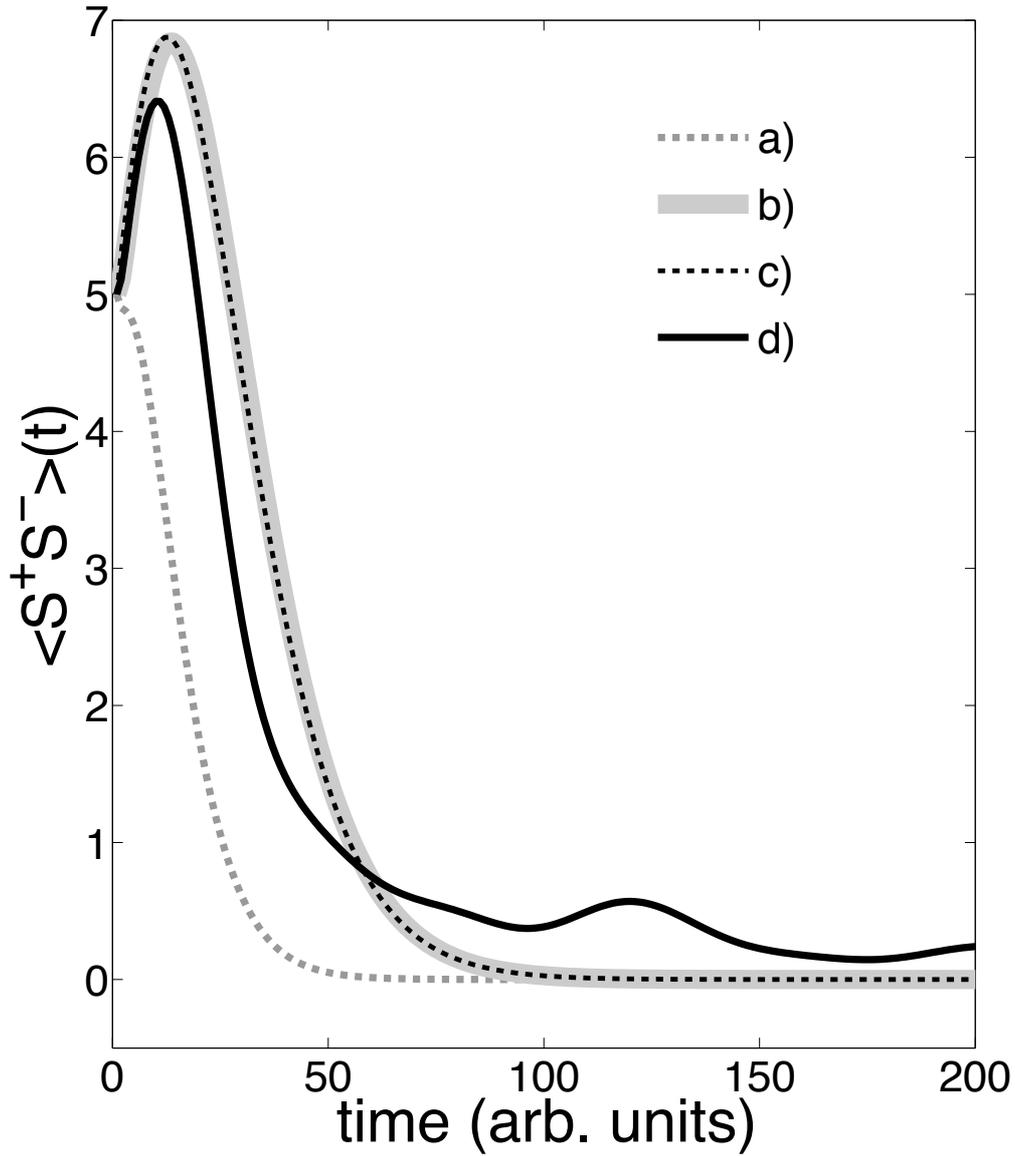

**Figure 2. Time dependence of the atomic intensity $\langle S^+ S^- \rangle(t)$ with a cavity decay rate $\kappa = 1$, a coupling $\lambda = 0.1$, and no pump acting on the system.**
a) resonant configuration $\omega_F = \omega_{A,i}$ and an atomic decay rate $\gamma = 0.02$.
b) resonant configuration and $\gamma = 0$.
c) superradiant result.
d) non resonant configuration with the following detuning with respect to the cavity frequency $\omega_F$:
$\Delta\omega_{A,1} = -0.01, \Delta\omega_{A,2} = 0.05, \Delta\omega_{A,3} = -0.07, \Delta\omega_{A,4} = 0.09, \Delta\omega_{A,5} = 0.03,$ **and $\gamma = 0$.**

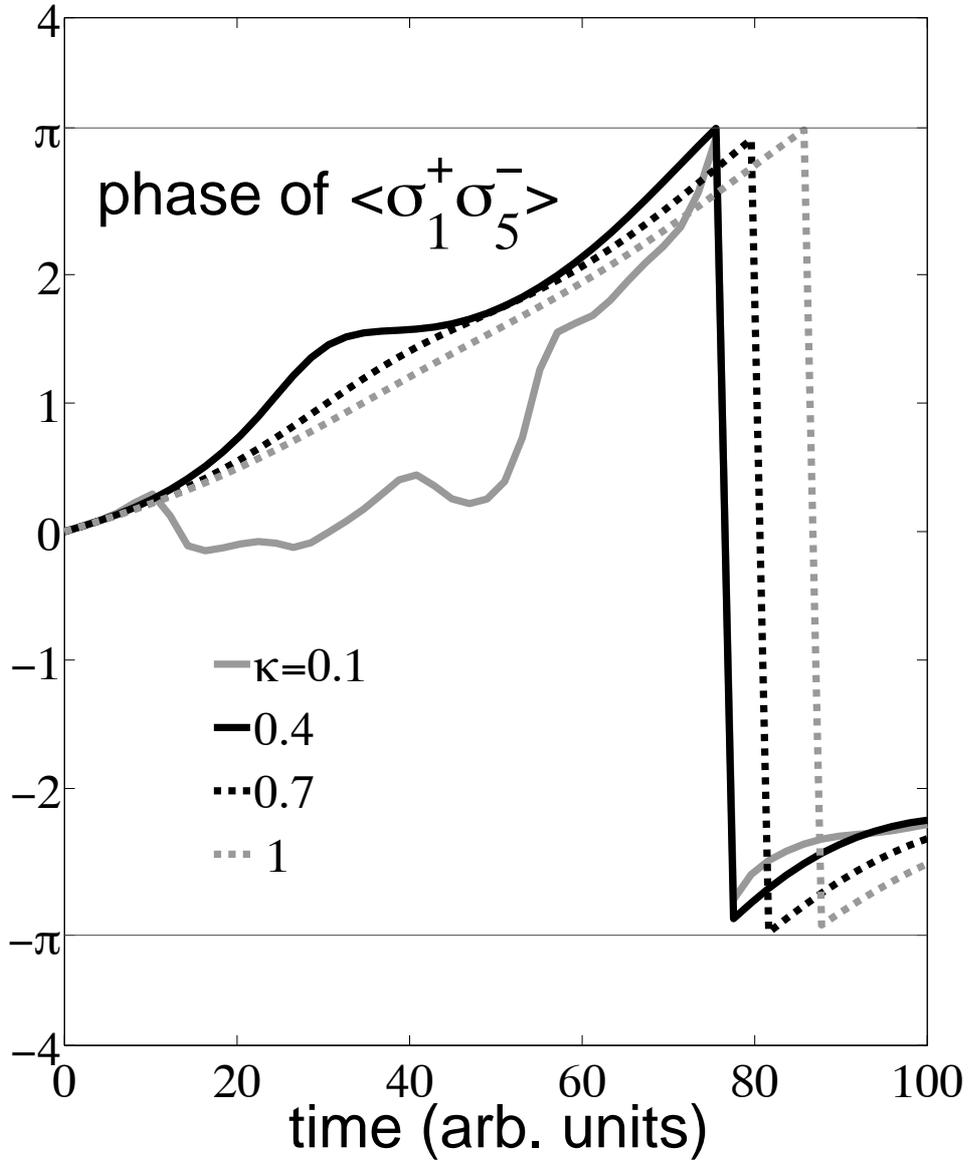

**Figure 3. Plot of the phase of $\langle \sigma_1^+ \sigma_5^- \rangle(t)$ as a function of time in a non-resonant configuration with the following detuning with respect to the cavity frequency $\omega_F$: $\Delta\omega_{A1} = -0.01$, $\Delta\omega_{A2} = 0.05$, $\Delta\omega_{A3} = -0.07$, $\Delta\omega_{A4} = 0.09$, $\Delta\omega_{A5} = 0.03$. The decay $\kappa$ takes various values, $\gamma = 0$, and $\lambda = 0.1$.**

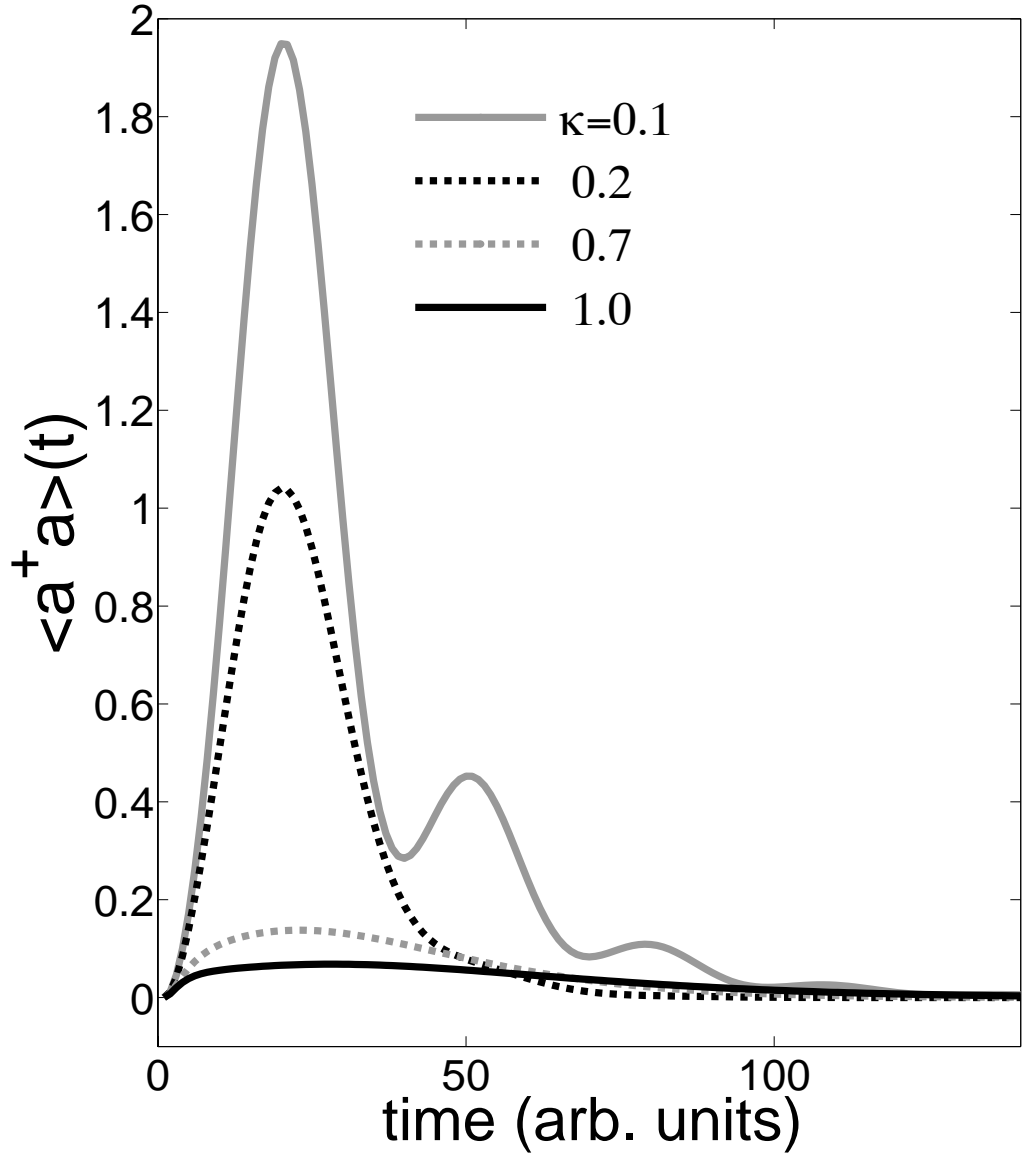

**Figure 4. Time dependence of the average photon number $\langle a^+a \rangle(t)$ for the resonant case $\omega_F = \omega_A$ and for various values of the cavity decay rate $\kappa$, with an atomic decay rate $\gamma = 0$, and a coupling $\lambda = 0.1$.**

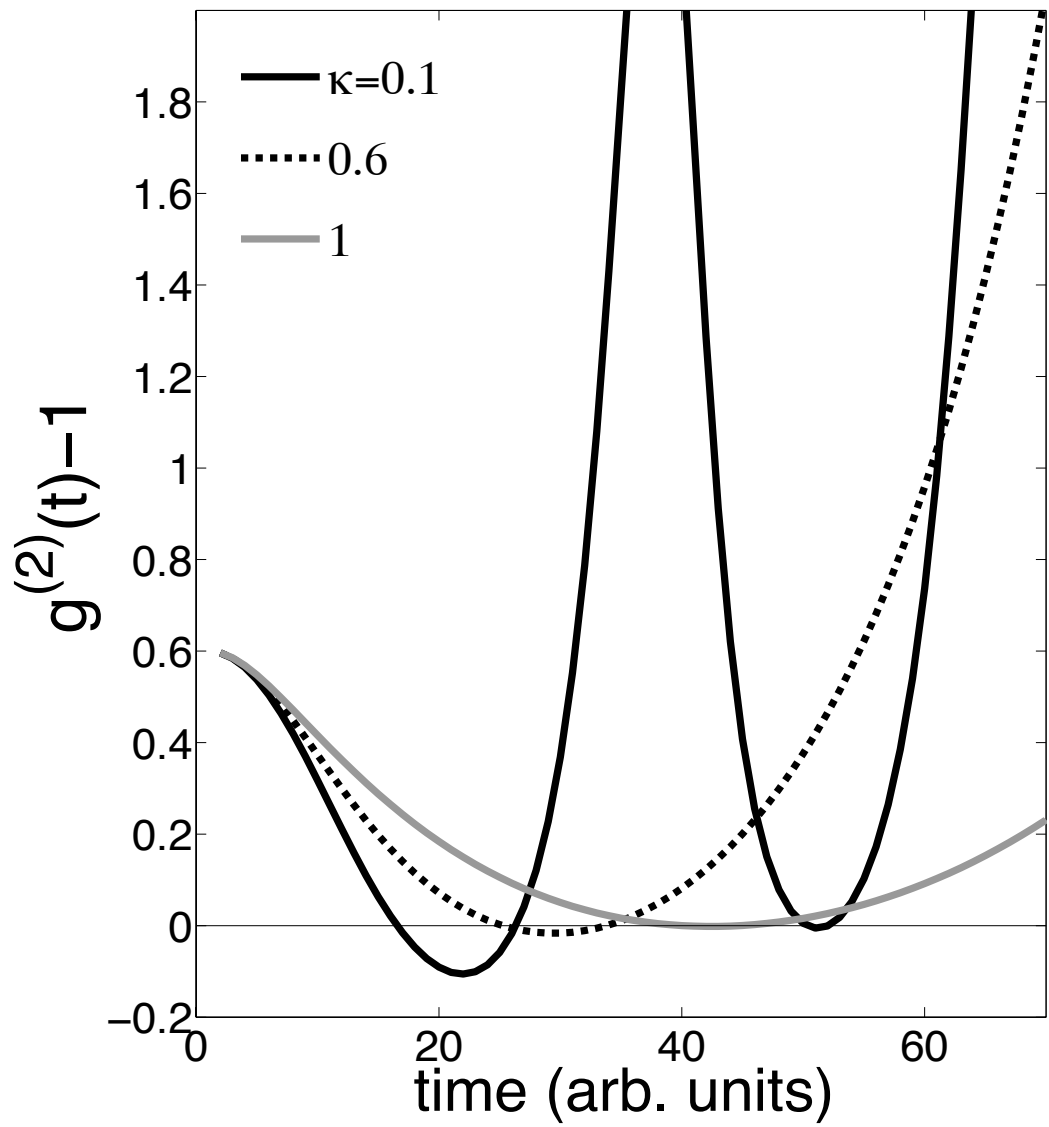

**Figure 5.** Plot of the photon second order correlation $g^{(2)}(t)-1$ as a function of time for various values of $\kappa$, $\gamma = 0$, and $\lambda = 0.1$.

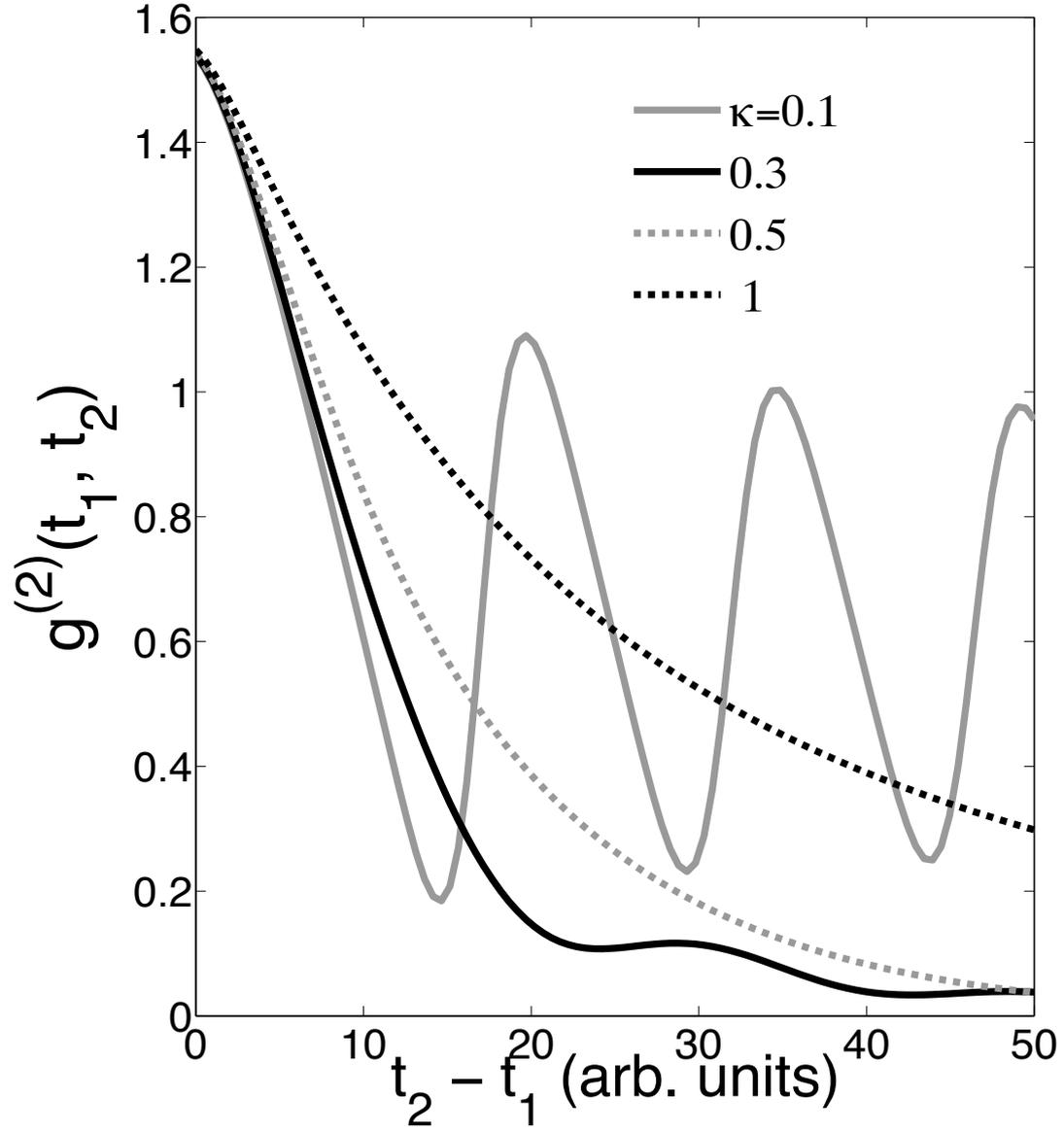

**Figure 6. Plot of the two-time-dependent normalized second order correlation** $g^{(2)}(t_1, t_2)$ **for** $t_1 = 2$ **units and** $t_2 \geq t_1$ **in the resonant configuration, for various values of** $\kappa$**, for** $\gamma = 0$**, and** $\lambda = 0.1$**.**

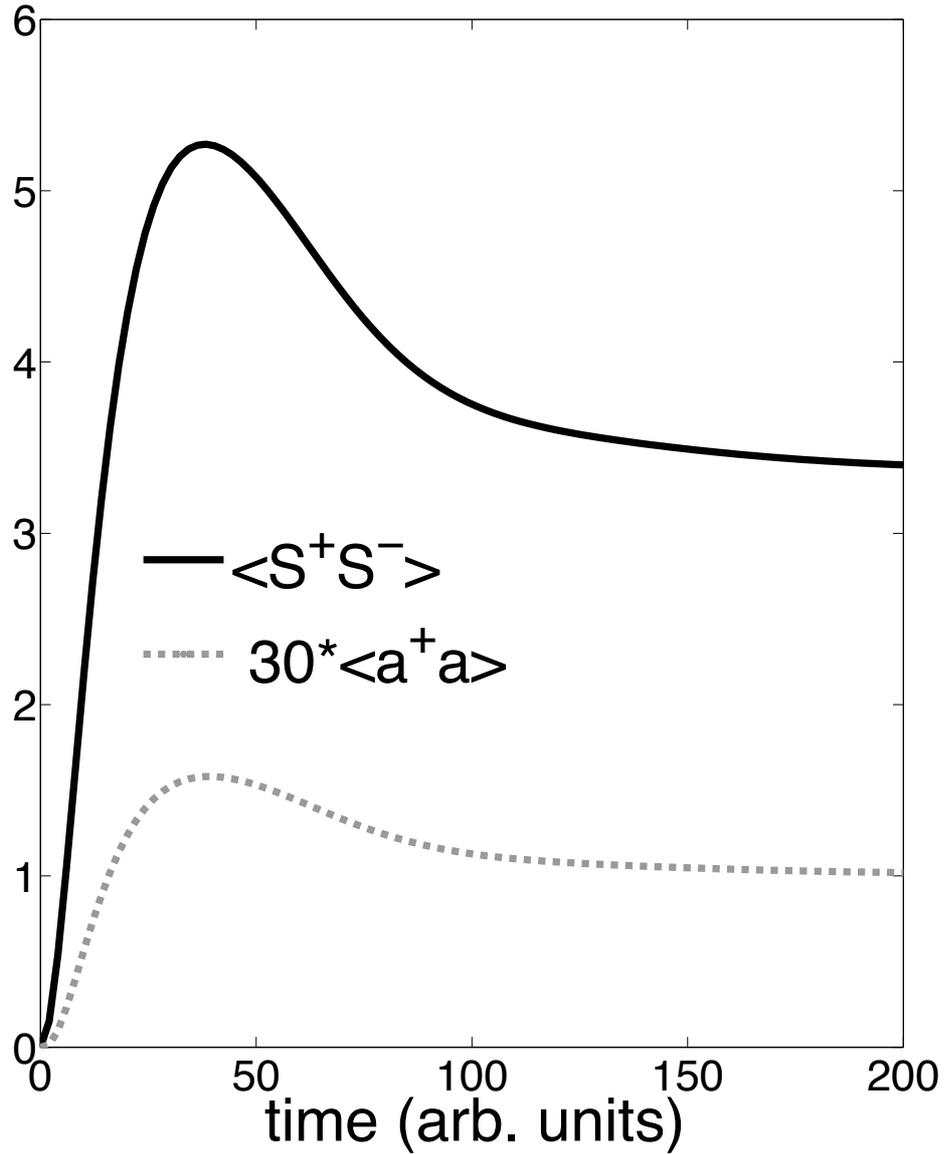

**Figure 7.** Plot of the atomic intensity $\langle S^+ S^- \rangle(t)$ and field intensity $\langle a^+ a \rangle(t)$ as a function of time for five 2-level atoms in a one-mode cavity, $N_{max} = 5$. At time t=0 the system is in the ground state. The parameters are $\kappa = 1$, $\gamma = 0$, and $\lambda = 0.1$, with a pump intensity $\eta = 0.04$ acting on the atoms.

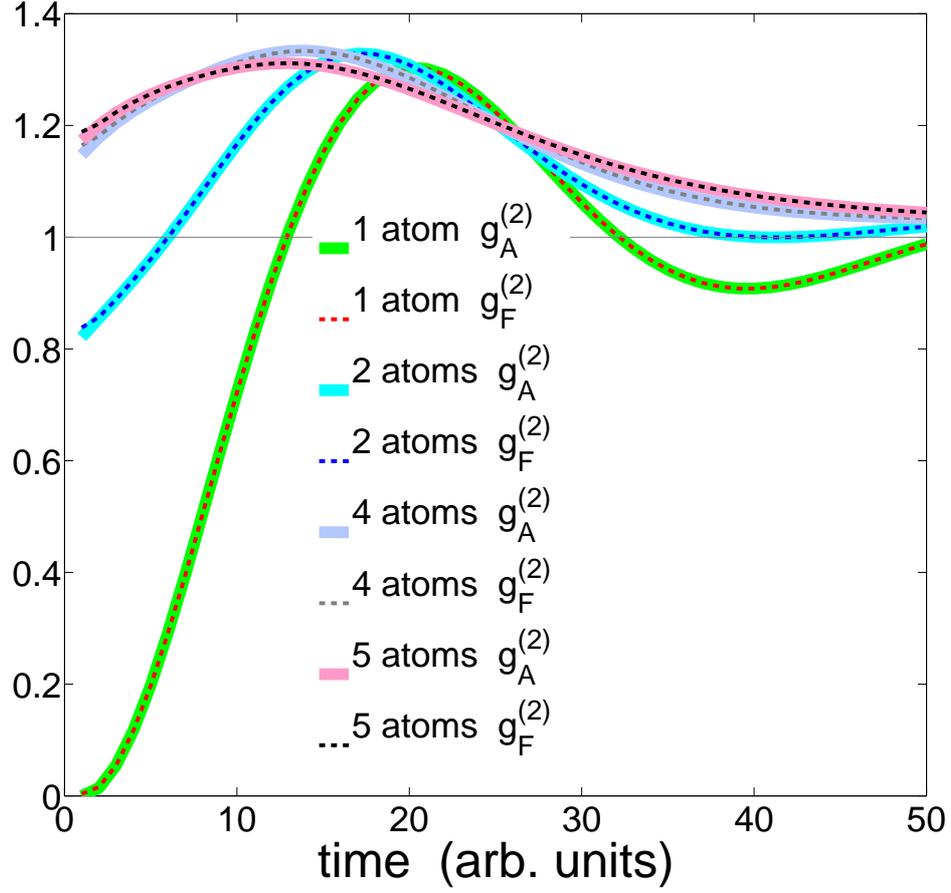

**Figure 8. Plot of $g_F^{(2)}(t)$ and $g_A^{(2)}(t)$ as a function of time for various number of atoms in a one mode bad cavity. The parameters are $\kappa = 1$, $\gamma = 0.01$, $\lambda = 0.1$ with a pump intensity $\eta = 0.04$ acting on the atoms.**

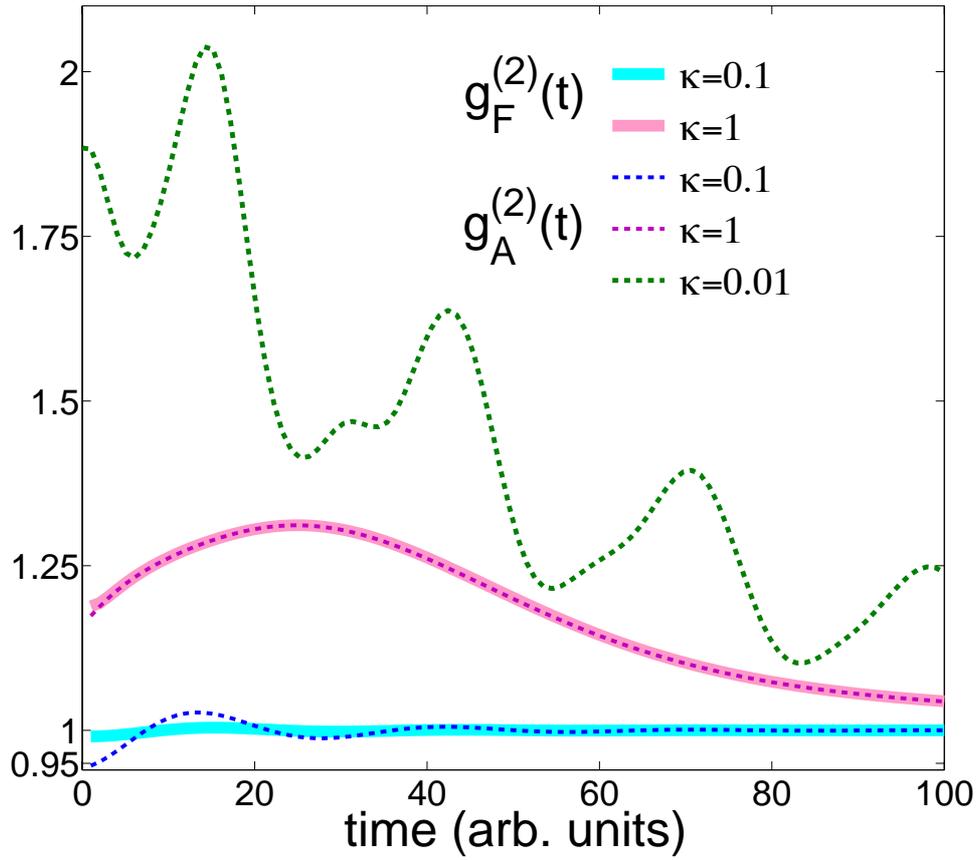

**Figure 9.** Plot of $g_F^{(2)}(t)$ and $g_A^{(2)}(t)$ as a function of time for five 2-level atoms in a one-mode cavity, $N_{max} = 5$. The parameters are $\gamma = 0.01$, $\lambda = 0.1$ with a pump intensity $\eta = 0.04$ acting on the atoms. The correlation $g_F^{(2)}(t)$ for $\kappa = 0.01$ shows very small oscillations around the value 1 and is not reported here.

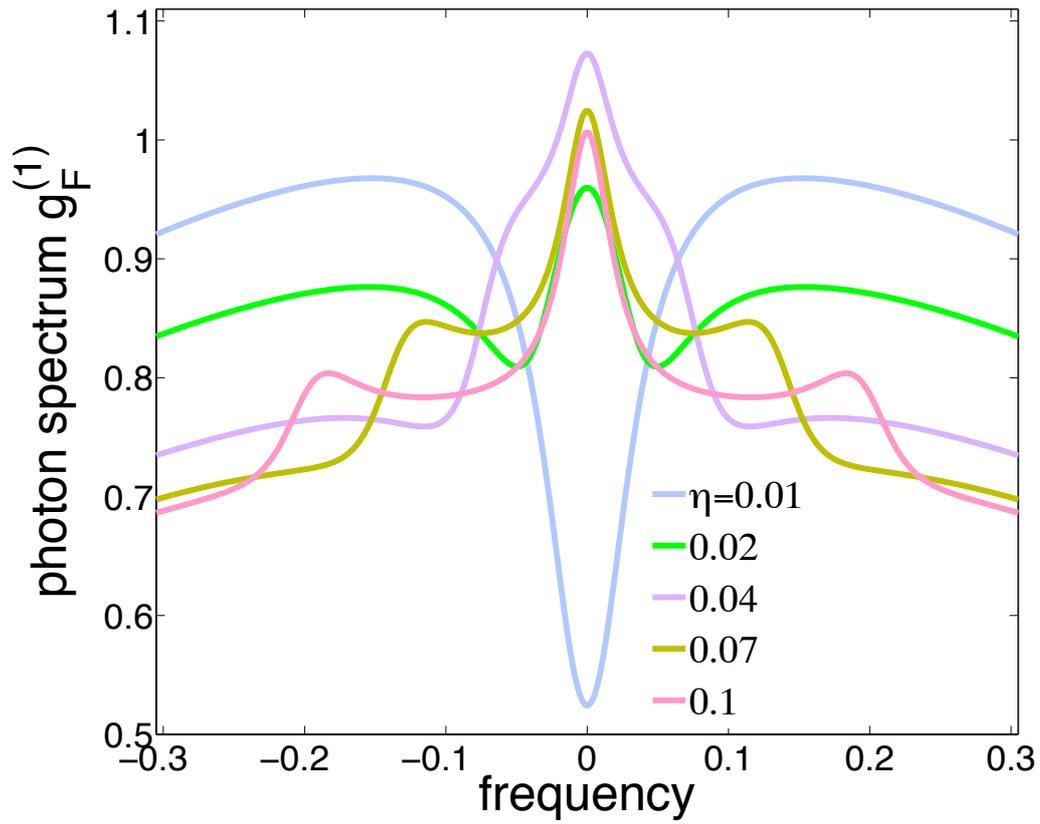

**Figure 10. Plot of the photon spectrum $g_F^{(1)}(\omega)$ as a function of the frequency $\omega$ for two 2-level atoms in a one-mode cavity, $N_{max} = 4$ for $\kappa = 1$, $\gamma = 0.01$, $\lambda = 0.1$, and for various values of the pump intensity $\eta$ acting on the atoms.**

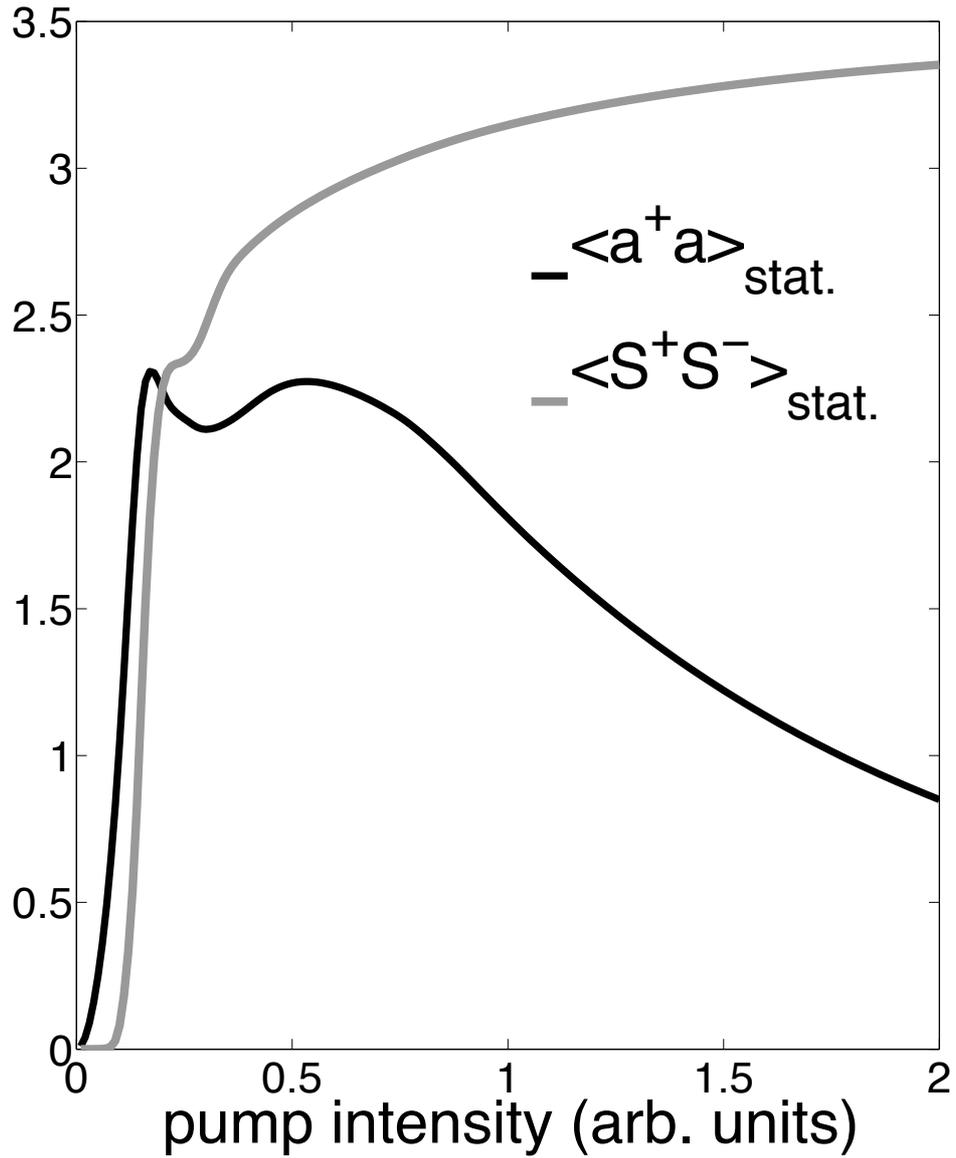

**Figure 11. Plot in the resonant configuration of the number of photons $\langle a^+ a \rangle_{stat.}$ in the cavity mode and of the atomic intensity $\langle S^+ S^- \rangle_{stat.}$ in the stationary regime as a function of the pump intensity $\alpha$ acting on the field mode. Four 2-level atoms are embedded in an unique cavity, $N_{max} = 7$. The cavity decay rate $\kappa = 0.01$, the atomic decay rate $\gamma = 0.01$ and the coupling $\lambda = 0.1$.**

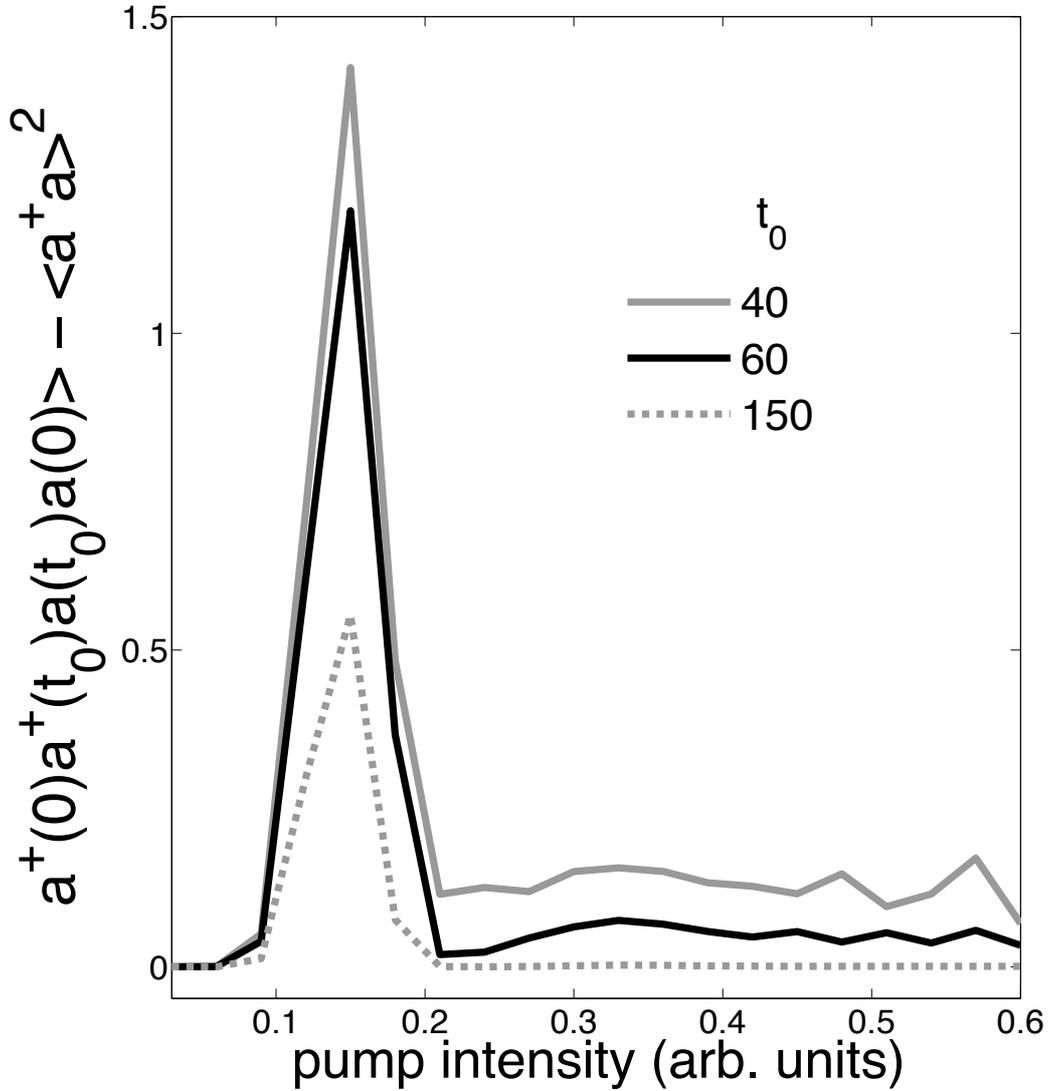

**Figure 12. Plot in the resonant configuration of the photon stationary fluctuations** $\left\langle a^+(0)a^+(t_0)a(t_0)a(0)\right\rangle_{stat.} - \left\langle a^+a\right\rangle^2_{stat.}$ **for various fixed times** $t_0$ **as a function of the pump intensity** $\alpha$ **acting on the field mode. Four two-level systems are embedded in a unique cavity,** $N_{max} = 7$. **The cavity decay rate** $\kappa = 0.01$, **the atomic decay rate** $\gamma = 0.01$ **and the coupling** $\lambda = 0.1$.

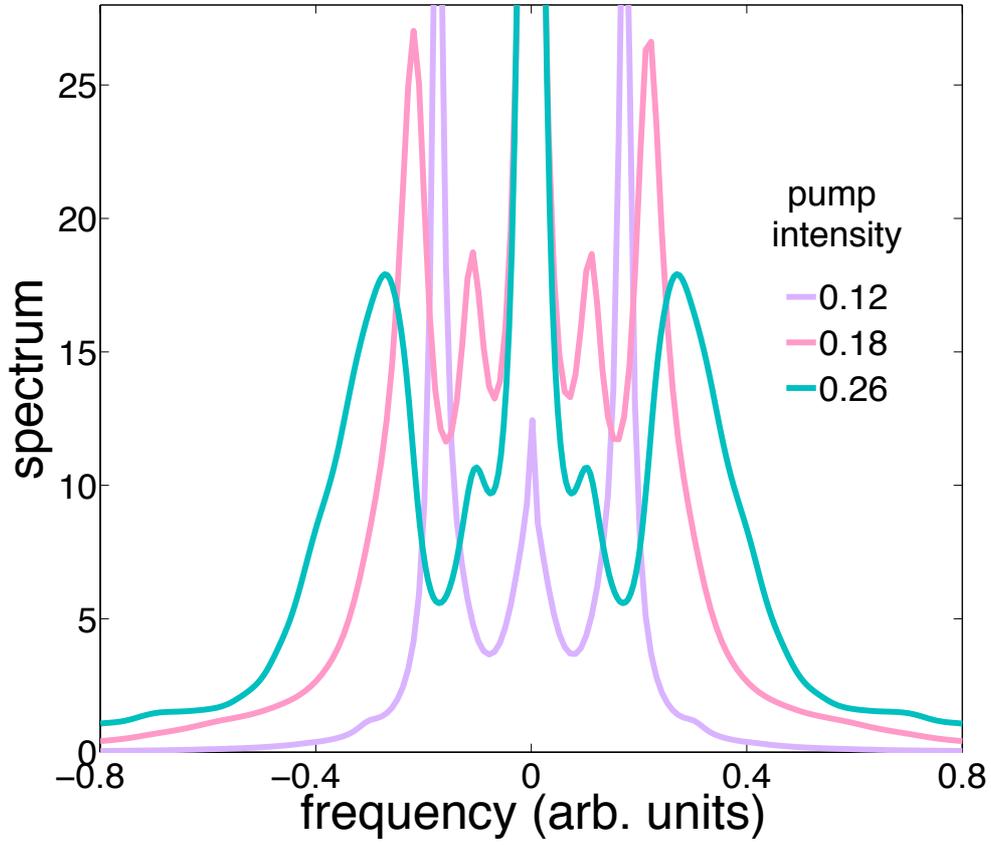

**Figure 13.** Plot in the resonant configuration of the photon spectrum $g_F^{(1)}(\omega)$ as a function of the frequency $\omega$ for $\kappa = \gamma = 0.01$, $\lambda = 0.1$, and for various values of the pump intensity $\alpha$ acting on the field mode. Four two-level systems are embedded in a unique cavity, $N_{max} = 7$.